%% file: main.tex
\documentclass[sigconf]{acmart}
\AtBeginDocument{%
  }

\usepackage{balance}
\usepackage{latexsym}
\usepackage[T1]{fontenc}
\usepackage[utf8]{inputenc}
\usepackage{microtype}
\usepackage{bigstrut}
\usepackage{boldline}
\usepackage{graphicx}
\usepackage{filecontents}
\usepackage{fancybox}
\usepackage{amsfonts}
\usepackage{amsmath}

\usepackage{amssymb}
\usepackage{bbm}
\usepackage{booktabs}
\usepackage{xspace}
\usepackage{cleveref}
\usepackage{enumitem}
\usepackage{graphicx}
\usepackage{multirow}
\usepackage{subcaption}
\usepackage{array}
\usepackage{makecell}
\usepackage{tikz}
\usepackage{mdframed}
\usetikzlibrary{shadows}
\usepackage{colortbl}
\usepackage{xcolor}
\usepackage{listings}
\usepackage{soul}
\usepackage{titlesec}

\titleformat{\subsection}
  {\normalfont\fontsize{10}{15}\bfseries}{\thesubsection}{1em}{}

\input{pygments.tex}
\input{defs}

\setcopyright{acmlicensed}
\copyrightyear{2018}
\acmYear{2018}
\acmDOI{XXXXXXX.XXXXXXX}
\acmConference[Conference acronym 'XX]{Make sure to enter the correct
  conference title from your rights confirmation email}{June 03--05,
  2018}{Woodstock, NY}
\acmISBN{978-1-4503-XXXX-X/2018/06}

\begin{document}

%\title{\sysname{}: Enhancing the Code Translation Ability of Language Models by Generating Multi-Modal Specifications}
\title{\tool{}: Automated Proof-of-Vulnerability Generation using LLM Agents}

\author{Vikram Nitin}
\authornote{Work done when the author was an intern at Microsoft}
\authornotemark[1]
\email{vikram.nitin@columbia.edu}
\affiliation{%
  \institution{Columbia University}
  \city{New York}
  \state{NY}
  \country{USA}
}

\author{Baishakhi Ray}
\email{rayb@cs.columbia.edu}
\affiliation{%
  \institution{Columbia University}
  \city{New York}
  \state{NY}
  \country{USA}
}

\author{Roshanak Zilouchian Moghaddam}
\email{rozilouc@microsoft.com}
\affiliation{%
  \institution{Microsoft}
  \city{Redmond}
  \state{WA}
  \country{USA}}

\renewcommand{\shortauthors}{Nitin et al.}

\input{sections/0_abstract}




%%z
\maketitle

\input{sections/1_introduction}
\input{sections/2_motivation}
\input{sections/3_methodology}
\input{sections/4_setup}
\input{sections/5_results}
\input{sections/6_related}
\input{sections/7_threats}
\input{sections/8_conclusion}

\bibliographystyle{ACM-Reference-Format}
\bibliography{references}
\clearpage
\appendix
\input{sections/supplementary}

\end{document}

%% file: pygments.tex
\newcommand\javadocblue[1]{\textcolor[rgb]{0.25,0.35,0.75}{{#1}}}

\definecolor{blueish}{RGB}{250, 250, 255}
\definecolor{greenish}{RGB}{250, 255, 250}
\definecolor{redish}{RGB}{255, 200, 200}

\definecolor{highlight}{RGB}{175, 255, 100}
\definecolor{gray01}{gray}{.98}
\definecolor{gray05}{gray}{0.95}
\definecolor{gray08}{gray}{0.92}
\definecolor{gray10}{gray}{0.90}
\definecolor{gray12}{gray}{0.88}
\definecolor{gray15}{gray}{0.85}
\definecolor{gray18}{gray}{0.82}
\definecolor{gray20}{gray}{0.80}
\definecolor{gray25}{gray}{0.75}
\definecolor{gray30}{gray}{0.70}
\definecolor{gray35}{gray}{0.65}
\definecolor{gray40}{gray}{0.60}
\definecolor{gray45}{gray}{0.55}
\definecolor{gray50}{gray}{0.50}
\definecolor{gray55}{gray}{0.45}
\definecolor{gray60}{gray}{0.40}
\definecolor{gray65}{gray}{0.35}
\definecolor{gray70}{gray}{0.30}
\definecolor{gray75}{gray}{0.25}
\definecolor{gray80}{gray}{0.20}
\definecolor{gray85}{gray}{0.15}
\definecolor{gray90}{gray}{0.10}
\definecolor{gray95}{gray}{0.05}

\definecolor{dkgreen}{RGB}{0,153,0}

%% file: defs.tex
\usepackage{enumitem}

\newcommand{\bi}{\begin{itemize}[leftmargin=*, wide=0pt, itemsep=0pt, parsep=0pt]}
\newcommand{\ei}{\end{itemize}}
\newcommand{\beq}{\begin{equation}}
\newcommand{\eeq}{\end{equation}}
\newcommand{\be}{\begin{enumerate}[leftmargin=*, wide=0pt, itemsep=0pt, parsep=0pt]}
\newcommand{\ee}{\end{enumerate}}

\newcommand{\tool}{\textsc{FaultLine}\xspace}

\newcommand\res[1]{}

\newcommand{\eg}{\hbox{\emph{e.g., }}\xspace}
\newcommand{\ie}{\hbox{\emph{i.e., }}\xspace}

\definecolor{javared}{rgb}{0.6,0,0} % for strings
\definecolor{javagreen}{rgb}{0.25,0.5,0.35} % comments
\definecolor{javapurple}{rgb}{0.5,0,0.35} % keywords
\definecolor{javadocblue}{rgb}{0.25,0.35,0.75} % javadoc

\newcommand{\TT}[1]{{\javadocblue{\small\texttt{#1}}}}

\definecolor{bluekeywords}{rgb}{0,0,1}
\definecolor{greencomments}{rgb}{0,0.5,0}
\definecolor{greennumbers}{rgb}{0,0.5,0}
\definecolor{redstrings}{rgb}{0.64,0.08,0.08}
\definecolor{blueident}{rgb}{0, 0.08, 0.45}

\lstdefinestyle{myc}{
  language=c,
  showstringspaces=false,
  basicstyle=\small\ttfamily,
  keywordstyle=\color{bluekeywords},
  commentstyle=\color{greencomments},
  identifierstyle=\color{blueident},
  stringstyle=\color{redstrings},
  frame=none,
  backgroundcolor=\color{white},
  escapeinside={<@}{@>},
  columns=fullflexible,
  breakatwhitespace=true,
  breaklines=true,
  literate=
  *{0}{{{\color{greennumbers}0}}}1
  {1}{{{\color{greennumbers}1}}}1
  {2}{{{\color{greennumbers}2}}}1
  {3}{{{\color{greennumbers}3}}}1
  {4}{{{\color{greennumbers}4}}}1
  {5}{{{\color{greennumbers}5}}}1
  {6}{{{\color{greennumbers}6}}}1
  {7}{{{\color{greennumbers}7}}}1
  {8}{{{\color{greennumbers}8}}}1
  {9}{{{\color{greennumbers}9}}}1
}

\definecolor{yellowhl}{rgb}{1,0.91,0.75}
\sethlcolor{yellowhl}

\titlespacing\subsection{0pt}{6pt plus 4pt minus 2pt}{0pt plus 2pt minus 2pt}

\newcommand{\result}[1]{\noindent\shadowbox{\parbox{0.96\linewidth}{\textbf{Summary:}~#1}}}

%% file: sections/0_abstract.tex
\begin{abstract}
Despite the critical threat posed by software security vulnerabilities, reports are often incomplete—lacking the proof-of-vulnerability (PoV) tests needed to validate fixes and prevent regressions. These tests are crucial not only for ensuring patches work, but also for helping developers understand exactly how vulnerabilities can be exploited. Generating PoV tests is a challenging problem, requiring reasoning about the flow of control and data through deeply nested levels of a program.

We present \tool{}, an LLM agent workflow that uses a set of carefully designed reasoning steps, inspired by aspects of traditional static and dynamic program analysis, to automatically generate PoV test cases. Given a software project with an accompanying vulnerability report, \tool{} 1) traces the flow of an input from an externally accessible API (``source'') to the ``sink'' corresponding to the vulnerability, 2) reasons about the conditions that an input must satisfy in order to traverse the branch conditions encountered along the flow, and 3) uses this reasoning to generate a PoV test case in a feedback-driven loop. \tool{} does not use language-specific static or dynamic analysis components, which enables it to be used across programming languages.

To evaluate \tool{}, we collate a challenging multi-lingual dataset of 100 known vulnerabilities in Java, C and C++ projects. On this dataset, \tool{} is able to generate PoV tests for 16 projects, compared to just 9 for CodeAct 2.1, a popular state-of-the-art open-source agentic framework. Thus, \tool{} represents a \textbf{77\%} relative improvement over the state of the art. Our findings suggest that hierarchical reasoning can enhance the performance of LLM agents on PoV test generation, but the problem in general remains challenging even for state-of-the-art models. We make our code and dataset publicly available in the hope that it will spur further research in this area.\footnote{\url{https://github.com/faultline-pov/icse-26}}
\end{abstract}

%% file: sections/1_introduction.tex
\section{Introduction}
\label{sec:intro}

Security vulnerabilities pose a significant threat to the software development process, driving the community to build various automated detection and fixing tools \cite{harzevili2023survey}. When a vulnerability is detected in a project, it is reported to the developers of the project, who then attempt to quickly fix it. Subsequently, a report is generated in the National Vulnerability Database (NVD) \cite{nist_nvd_2024}, containing a textual description of the vulnerability and mitigation strategies for users of the software. However, most of these reports~\res{generated by static analyzers? as the dynamic analysis will have inputs, isn't it?} lack Proof-of-Vulnerability (PoV) tests that demonstrate the vulnerability. PoV tests are designed to \textit{fail} when the vulnerability exists, and \textit{succeed} when the vulnerability is fixed. Thus, they act as an oracle to verify the effectiveness of the fix, and ensure that the vulnerability is not inadvertently reintroduced during future development of the project. In addition, they can enable developers to better understand the vulnerability. Studies have shown~\cite{mu2018understanding} that human developers struggle to reproduce vulnerabilities from reports, because these reports frequently miss crucial information. PoV tests complement the information in a report and provide a clear demonstration of the exploit.

\begin{figure*}[t]
    \includegraphics[width=0.9\linewidth]{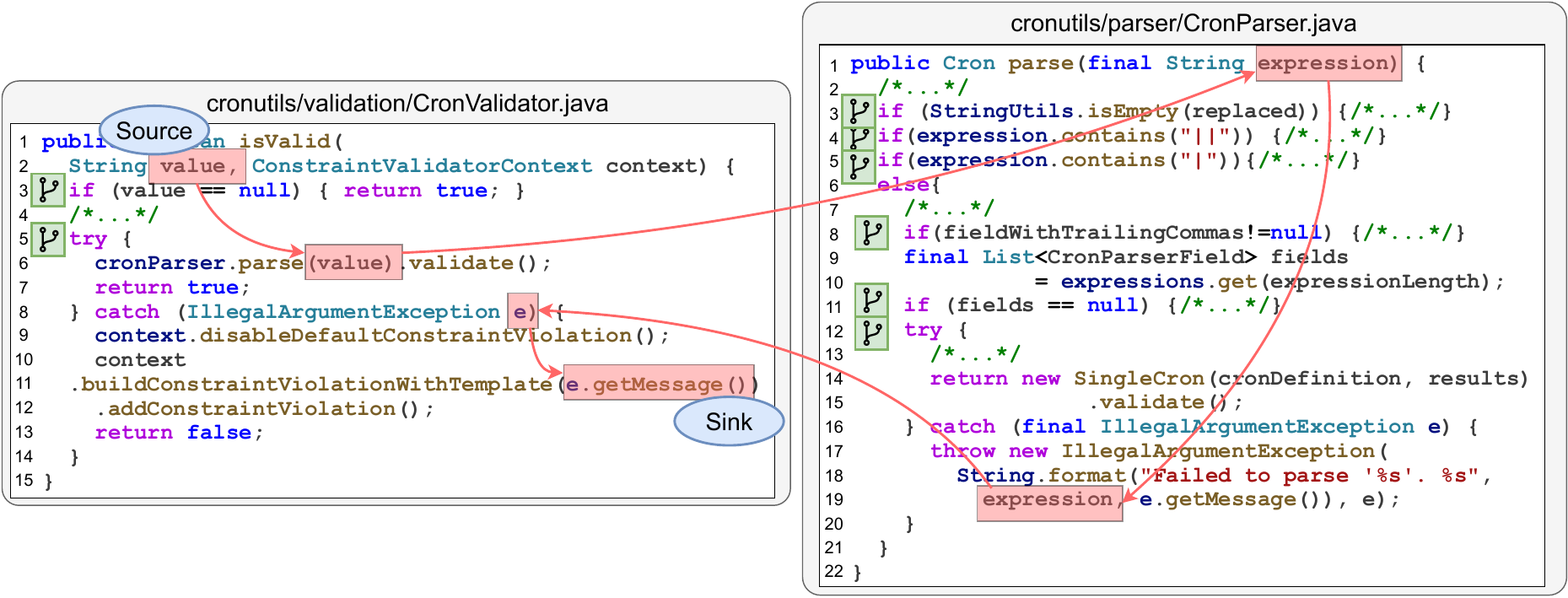}
    \vspace{4pt}
    \caption{Motivating example - A Code Injection vulnerability from the \TT{cron-utils} Java library. The vulnerable flow is highlighted in red, and the branch conditions are marked in green. The String \TT{value} is taken as a user-provided argument on line 2 of \TT{CronValidator.java}, and it flows to the Sink on line 11 where any embedded code could potentially be executed. \res{Explain the vulnerability here}}
    \label{fig:motivating}
\end{figure*}

\noindent
\textbf{Existing work and limitations:} Recently, Large Language Models (LLMs) have been used as components in \textit{autonomous agents} to solve various software engineering tasks~\cite{yang2024swe, wang2025openhands, zhang2024autocoderover}. These systems augment LLMs with the ability to invoke \textit{tools} to read, write and execute code, enabling the LLM to interact with the code base much as a human developer would. However, constructing PoV tests is a challenging problem for LLM agents~\cite{zhu2025cve}. Some of the reasons for this are:
% constructing PoV tests is a
% challenging problem for LLM agents for two main reasons: 1) LLMs are not usually trained with data flow or control flow information, and therefore they do not leverage these information too solve programming tasks. 2) LLMs are not yet fully follow all instructions and satisfy all input conditions, while a PoV test needs to satisfy multiple conditions, it needs to demonstrate the vulnerability, fails when the vulnerability exists, and pass when it is resolved.
\bi
\item \textit{Insufficient understanding of data flow.} A vulnerability exploit starts with an externally accessible API or user input (``source''), and traverses through multiple function calls until it reaches the location where the vulnerability occurs (``sink''). Writing a PoV test involves reasoning about this flow, and invoking the precise methods that trigger it. LLMs are usually not trained on data flow traces, and therefore they do not effectively leverage this type of reasoning to solve programming tasks.~\res{How do you know about such LLM behavior? Any proof?}.

\item \textit{Insufficient understanding of control flow.} The program path from source to sink frequently involves many branch conditions, that divert the flow into paths that do not reach the vulnerability. The input in a PoV test must be carefully crafted such that the program flow proceeds along the correct path at each branch. LLM agents often miss certain crucial conditions on test inputs, and the tests do not reach the vulnerability. Additionally, they are unable to systematically reason about the cause of this failure and refine the test.
\item \textit{Misalignment with initial goals.} A PoV test must satisfy certain requirements - it must fail when the vulnerability exists, and demonstrate the exploit by actually running the vulnerable code. LLM agents frequently stop after generating a test that satisfies \textit{some} (but not all) of these requirements. For instance, they may generate a test that simply reads the source code to check for the presence of a particular line of code corresponding to the vulnerability, and the test does not actually build the project or run the vulnerable code.

\res{May be also talk about entry point than just unit test cases}
\ei

\noindent
\textbf{Our approach:} To address the above shortcomings, we propose \tool{}, a workflow-based LLM agent that uses a composition of carefully designed reasoning steps to design a PoV test for a known vulnerability. Unlike existing agents that generate tests with an incomplete or incorrect understanding of program semantics, \tool{} prompts an LLM to extract certain semantic properties of the program~\res{what kind of semantic information?} before generating a test. Specifically, given a program along with an accompanying vulnerability report, \tool{} traces the flow of data from source to sink, reasons about the requirements that a test must satisfy in order to cover this path, uses these insights to generate a PoV test, and refines it in a feedback-guided loop.

\noindent
\textbf{Results summary:} To evaluate \tool{}, we collate a challenging multi-lingual dataset of 100 known vulnerabilities in Java, C and C++ projects. Our key findings are listed below:
\bi
\item On this dataset, \tool{} is able to generate correct PoV tests for 16 projects. In comparison, CodeAct 2.1, a popular agentic framework, is able to generate correct PoV tests for only 9.
\item The tests generated by \tool{} reach the functions or methods corresponding to the vulnerability for 31 projects, compared to 21 for the baseline.
\item We show that both flow reasoning and branch reasoning are essential to \tool{}'s performance.
\ei

\noindent
\textbf{Contributions:} This paper makes the following contributions to the state of the art:
\be
\item We design an agentic workflow based on a series of carefully crafted reasoning steps to generate PoV tests.
\item We empirically establish the effectiveness of this workflow in generating PoV tests, and highlight the importance of each component of the workflow.
\item We contribute a benchmark for PoV test generation, comprising 100 vulnerabilities spanning 4 CWE categories. This dataset challenges LLMs to reason about extremely subtle properties of a program, and represents a frontier for LLM-based code reasoning.
\ee

%% file: sections/2_motivation.tex
\section{Background and Motivating Example}

\subsection{Background}

A security vulnerability in a software project manifests as a \textit{flow} that leads from a ``source'' to a ``sink''. A \textbf{source} is traditionally defined as a program point where data enter the program from external or untrusted sources. Some examples are external API functions (for software libraries), user form inputs (for web applications), or HTTP endpoints (for web services). More generally, any property of the program that can be controlled by an attacker can be considered a source. A \textbf{sink} is any program construct that can cause undesirable effects if attacker-controlled data is passed directly to it. For instance, a function that executes SQL queries is considered a sink because it can be invoked with SQL queries that delete or alter data stored in the linked database.

The fundamental principle of secure software design is to ensure that each flow between a source and a sink is properly filtered, or \textbf{sanitized}. Consider a web form that accepts text input from a user, and uses this text to retrieve matching records from a database. A quintessential vulnerability pattern in such a setting is \textit{SQL injection}, whereby an attacker embeds carefully crafted SQL queries in their input text. If this input text reaches a function that interacts with the database, the embedded SQL queries might be executed on the underlying database, leading to data loss or privacy concerns. To safeguard against this, developers must sanitize text inputs by checking for patterns that are indicative of SQL injection exploits.

Software vulnerabilities can be organized into groups depending on the nature of the sources and sinks. Common Weakness Enumeration (CWE) \cite{cwe_mitre} is a widely adopted categorization system for vulnerabilities. Each category is assigned a number, \eg CWE-94 corresponds to \textit{Code Injection} vulnerabilities, such as SQL injection. When a vulnerability is reported, it is usually assigned one or more CWE categories.

\begin{figure*}
\includegraphics[width=0.95\linewidth]{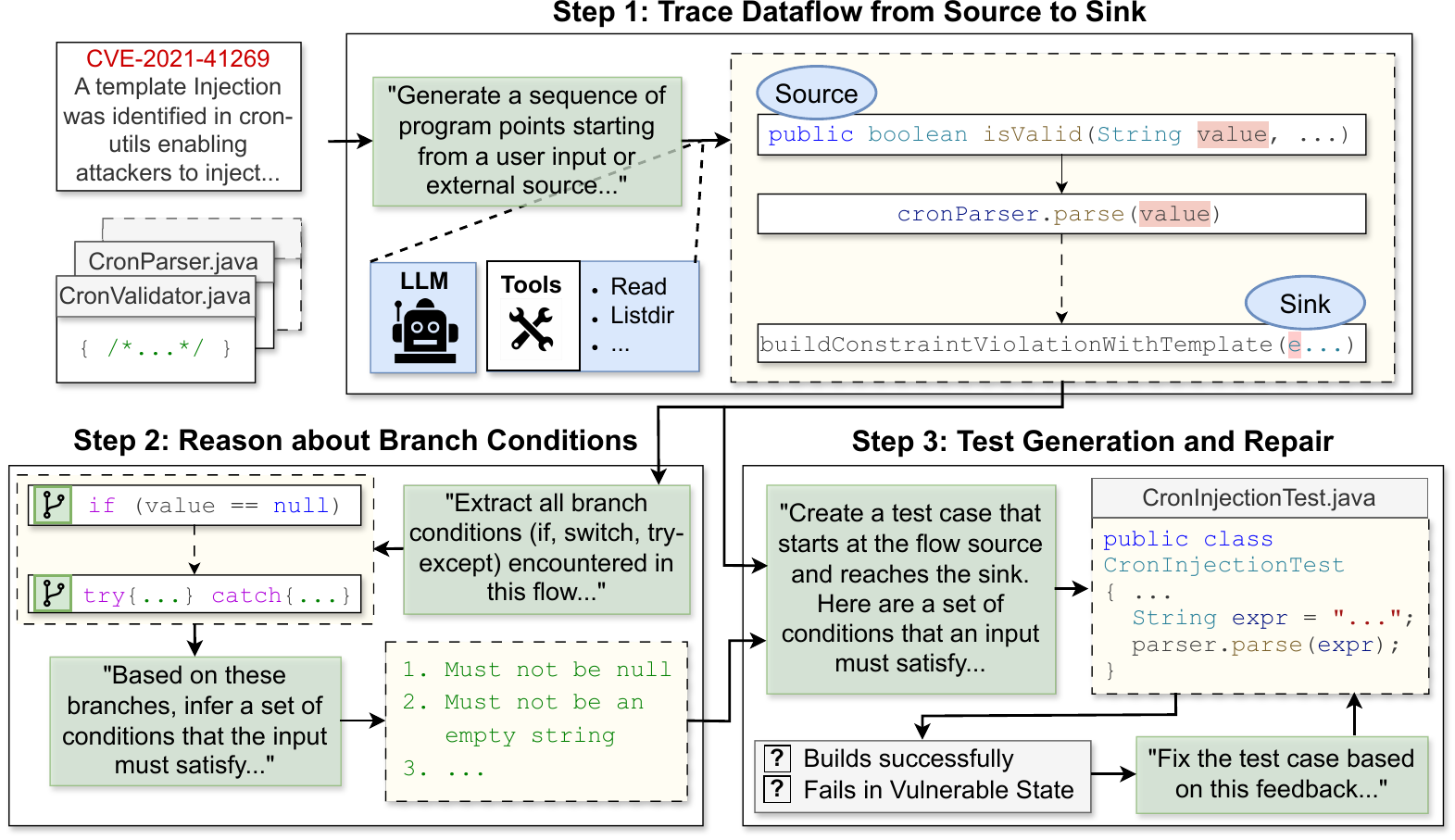}
\vspace{6pt}
\caption{Our system \tool{}, for generating Proof of Vulnerability test cases.}
\label{fig:sysdiagram}
\end{figure*}

\subsection{Motivation}
\label{sec:motivation}
In this section, we use a real-world example to highlight the challenges involved in generating PoV test cases. \Cref{fig:motivating} shows a \textit{Code Injection} vulnerability in the Java library \TT{cron-utils}. An attacker can exploit this to execute arbitrary code on the host system. When one tries to generate a PoV for this vulnerability, certain challenges arise:

\noindent
\textbf{Challenge 1:} \textit{Tracing the flow of data from source to sink.} A prerequisite for generating a PoV test is identifying a source and sink with an un-sanitized flow between them. In \Cref{fig:motivating}, the function \TT{isValid} can be called with a user-provided String \TT{value}, so this is a \textit{source}. The function \TT{buildConstraintViolationWithTemplate} on line 11 of \TT{CronValidator.java} accepts arguments written in Java's Expression Language (EL), and can execute arbitrary code. So this is a \textit{sink}. When we look at \TT{CronValidator.java}, it may seem like there is no flow between \TT{value} on line 2 and \TT{e} on line 11. However, when we look closer, we find that \TT{cronParser.parse()} on line 6 of \TT{CronValidator.java} throws an exception which is caught on Line 8. The error message includes the \TT{value} string verbatim, and this reaches the sink on line 11 with no sanitization. Tracing such a flow across multiple files is a challenge for both humans and LLM agents.

\noindent
\textbf{Challenge 2:} \textit{Crafting an input that circumvents the branch conditions.} Even though there may be a \textit{data flow} path from source to sink, this is only one of several execution paths that the program can take, depending on the \textit{control flow}. In order to exercise this particular path, we have to call the \TT{isValid} method with appropriate arguments such that the \textit{control flow} reaches line 11 of \TT{CronValidator.java}. The branch conditions are marked in green in \Cref{fig:motivating}. We can see that there are 6 \TT{if} conditions and 2 \TT{try..except} blocks. Each of these corresponds to a constraint on the input that has to be satisfied in order for the control flow to reach the vulnerability sink. For example, the condition on lines 9-11 of \TT{CronParser.java} expects the expression to have a certain number of fields (number of space-separated components). If there is a mismatch between \TT{expressionLength} and the expected number of fields, \TT{fields} will be \TT{null}, and the control flow will be diverted down a non-vulnerable path. Note that this figure only captures a portion of the full complexity of the program's control flow; for example, the \TT{validate()} function on line 15 of \TT{CronParser.java} also needs to fail and throw an \TT{IllegalArgumentException}. This means that an input that triggers this vulnerability needs to be extremely carefully crafted to circumvent all of these branch conditions\footnote{\url{https://securitylab.github.com/advisories/GHSL-2020-212-cron-utils-ssti/}}. This is representative of vulnerability-triggering inputs in general, as these tend to arise only in very specific edge cases.

\noindent
\textbf{Limitations of existing LLM agents:} LLM agents tend to struggle with this kind of complex reasoning, and perform poorly out of the box. For instance, when we use CodeAct 2.1, a popular LLM agent, ~\res{what is OpenHands? Give context} to generate a PoV test for the above vulnerability, it generates a test that calls \TT{isValid()} with the following string as the \TT{value} argument:\\ \TT{"\$\{new java.io.FileWriter('/tmp/...').write('exploit')\}"}\\
The intention is to inject Java code that writes to a file in the \TT{tmp} directory. However, the agent fails to understand that the expression has to have a \textit{specific number of fields} in order to get past the branch condition on lines 9-11 of \TT{CronParser.java}. A valid Cron expression needs to have \textbf{6} or \textbf{7} space-separated components, which is what that branch condition checks internally. The generated expression has only \textbf{2} space-separated components, namely \TT{"\$(new"} and \TT{"java.io...\}"}.

Our tool, \tool{}, uses a series of carefully crafted reasoning steps to guide LLM agents to address the above challenges and generate better PoV tests.

%% file: sections/3_methodology.tex
\section{Methodology}
\label{sec:methodology}
In this section, we describe \tool{}, an LLM agent that automatically generates PoV test cases for a project with a reported vulnerability. In this paper, we focus on the following vulnerability types:
\bi
\item \textit{CWE-22 (Path Traversal):} This occurs when insufficient validation of user-supplied input in file path construction allows attackers to access files outside the intended directory using sequences such as \TT{"../"}.
\item \textit{CWE-78 (OS Command Injection):} This enables attackers to execute arbitrary operating system commands by injecting malicious input into application-constructed system commands .
\item \textit{CWE-79 (Cross-Site Scripting):} This happens when unvalidated user input in web output allows attackers to execute malicious scripts in victims' browsers.
\item \textit{CWE-94 (Code Injection):} This allows attackers to inject and execute arbitrary code by exploiting insufficient input validation in code interpretation functions.
\ei
However, we emphasize \tool{} is not limited to these specific CWE categories, and our framework is general enough to permit extension to any other software vulnerability type.

\noindent
\Cref{fig:sysdiagram} describes the workflow of our agent. In the first stage (\Cref{sec:dataflow}), we prompt the LLM to identify a source and a sink with a flow between them. In the second stage (\Cref{sec:controlflow}), we leverage the agent to reason about the branch conditions encountered along the flow, and use these conditions to derive a set of conditions that an input has to satisfy in order to pass through this flow. In the final stage (\Cref{sec:testgen}), we use the information aggregated from previous stages to generate a PoV test case, and repair it based on build and execution feedback. Although the final reasoning generated in one stage is passed on to subsequent stages, the various stages do not share a common conversation memory. This keeps the length of each conversation in check. Full LLM prompts for all these stages are available in the appendix.

\subsection{Data Flow Reasoning}
\label{sec:dataflow}
We start with a project containing a known vulnerability, and the corresponding vulnerability report extracted from NVD database~\cite{nist_nvd_2024}. The first step towards generating a PoV test case is to process the report, scan the files in the repository, and understand the nature of the vulnerability. Often, the report provides scant information. This is sometimes done intentionally, so as to not inadvertently provide attackers with a blueprint on how to craft an exploit. However, every report includes at least a) the CWE categorization of the vulnerability, b) affected versions of the project, c) mitigation strategies for developers. \Cref{fig:report} shows an example of a vulnerability report corresponding to the motivating example in \Cref{sec:motivation}.

\begin{figure}
\begin{lstlisting}[basicstyle=\small\ttfamily\color{dkgreen}, breaklines=true, breakindent=0pt, escapeinside={(<}{>)}, columns=fullflexible]
(<\textbf{CWE-94}>): Improper Control of Generation of Code ('Code Injection')
(<\textbf{Details}:>) cron-utils is a Java library to define, parse, validate, migrate crons as well as get human readable descriptions for them. In affected versions A template Injection was identified in cron-utils enabling attackers to(<\hl{\mbox{inject arbitrary Java EL expressions}}>), leading to unauthenticated Remote Code Execution (RCE) vulnerability. Versions up to 9.1.2 are susceptible to this vulnerability. Please note, that only projects (<\hl{using the @Cron annotation \mbox{to validate untrusted Cron expressions}}>) are affected. The issue was patched and a new version was released. Please upgrade to version 9.1.6. There are no known workarounds.
\end{lstlisting}
\caption{An example of a vulnerability report, for CVE-2021-41269. Although it is vague, the highlighted sections provide some hints about the sources and sinks of this vulnerability.}
\label{fig:report}
\end{figure}

Although this report is vague, it provides hints that can be used to deconstruct the vulnerable flow. The report states that the vulnerability allows attackers to \hl{``inject arbitrary Java EL expressions''}. This narrows the search for \textit{sinks} to program constructs that can process Java EL expressions. Further, the report indicates that the vulnerability arises from \hl{``using the \mbox{\TT{@Cron}} annotation to validate untrusted Cron expressions''}. This means that the \textit{source} is likely to be an API that accepts a Cron expression from a user.

Given these semantic hints from the vulnerability report, the next step is to automatically identify the data flow from source to sink within the codebase. One natural approach would be to leverage existing static analysis tools designed for this purpose. CodeQL \cite{codeql}, for instance, is an industry standard tool used for detecting dangerous flows between sinks and sources. However, several limitations make it unsuitable for our use case: a) it cannot directly utilize semantic information from the report to guide its detection algorithm, b) it requires a non-trivial amount of effort to set up and run for each project and programming language, c) it is tuned for high precision, causing it to miss several flows. In fact, CodeQL fails to detect the flow corresponding to our motivating example \cite{li2025iris}!

For all of the above reasons, we opt to use an LLM acting as an autonomous agent to reconstruct the vulnerable flow. We include the entire vulnerability report in the initial prompt, and instruct the LLM agent to use the information in the report to generate a flow starting from a source and reaching a sink. To enable the agent to explore the project's source code, we provide it with tools to list a directory (\texttt{ListDir}) and read a file (\texttt{Read}). In order to use semantic hints from the vulnerability report, such as the \TT{@Cron} annotation, we additionally equip the agent with tools to search for files by name (\texttt{Find}), and to search for files containing specific strings (\texttt{Grep}).

The output of this stage is a sequence of program points comprising a flow, where each point is identified by a short snippet of code (1-2 lines). Each point is additionally labeled with:
\be
\item the name of the file it is contained in,
\item the name of the variable that carries the vulnerable flow, and
\item a role, \ie Source, Sink, or Intermediate Node. 
\ee
As an example, here is the portion of the data flow reasoning output corresponding to the \textit{source} of the flow in \Cref{fig:motivating}.
\begin{figure}[h!]
\begin{lstlisting}[basicstyle=\ttfamily, breaklines=true, breakindent=60pt, escapeinside={(*@}{@*)}]
{ "code": "public boolean isValid(String value, ...",
  "role": "Source",
  "variable": "value",
  "file": ".../CronValidator.java" }
\end{lstlisting}
\end{figure}

The entire flow output consists of a sequence of points in the above format, starting with a source and ending with a sink. We use this flow as an input for our subsequent reasoning and generation steps.

\subsection{Control Flow Reasoning}
\label{sec:controlflow}

Once a flow from source to sink has been identified, the next step towards generating a PoV test case is reasoning about how to generate an input that will actually traverse this flow. We decompose this task into two steps - extracting branch conditions, and reasoning about conditions on the input.

\noindent
\textbf{Extracting branch conditions.} We collect the flow reasoning generated by Step 1, and instruct an LLM agent to follow this flow and extract all the branch conditions that an input might encounter on the way. This includes not just \TT{if-else} and \TT{switch} constructs, but also \TT{try-except} blocks. Each branch represents a potential opportunity for the control flow to be diverted down a non-vulnerable path that never reaches the sink, or pass through program points that sanitize the flow and render it non-threatening. Reasoning about every single branch, therefore, is crucial to constructing an input that can reach the sink. We refer to the path represented by the program points corresponding to these branch conditions as the \textit{control flow path}.

Note that the control flow path can be \textit{very different} from the data flow path (\Cref{sec:dataflow}). For example, in \Cref{fig:motivating}, the data flow (shown in red) spans the two files \TT{CronValidator.java} and \TT{CronParser.java}. However, the control flow path includes a completely different set of nodes (shown in green). Further, a portion of the control flow path traverses a file not shown in this figure, \TT{SingleCron.java}. This is induced by the call to \\\TT{SingleCron::validate()} on line 14 of \TT{CronParser.java}, which must throw an \TT{\mbox{IllegalArgumentException}} in order for the program to proceed down the path towards the sink. So although one might imagine the data flow and control flow paths to be similar, they have many fundamental qualitative differences.

Similarly to \Cref{sec:dataflow}, we equip the LLM agent with \texttt{ListDir}, \texttt{Read}, \texttt{Find} and \texttt{Grep} tools. We prompt it to extract each branch condition as a short snippet of code (1-2 lines). Each condition is required to be additionally labeled with:
\be
\item its type (If-Else, Switch, etc.),
\item the name of the file it is contained in, and
\item the desired outcome of the branch, in order for the input to proceed down a path that leads to the sink.
\ee
For example, the portion of the branch reasoning output corresponding to the branch condition in line 3 of \TT{CronValidator.java} in \Cref{fig:motivating} is:
\begin{figure}[h!]
\begin{lstlisting}[basicstyle=\ttfamily, breaklines=true, breakindent=60pt, escapeinside={(*@}{@*)}]
{ "code": "if (value == null)",
  "type": "If-Else",
  "file": ".../CronValidator.java",
  "outcome": "False - the value should not be null" }
\end{lstlisting}
\end{figure}

\noindent
\textbf{Reasoning about conditions on the input.} Analyzing these branch conditions carefully can provide information on how to generate an appropriate test input that walks a metaphorical tightrope, traversing the correct path through these branches. However, we found that LLMs often struggled to parse these branch conditions and produced tests with inadequate inputs. To make the connection between branch conditions and input requirements more explicit, we add a further reasoning step.

We ask the agent to reflect on its own output and infer a set of conditions that an external input must satisfy, in order to pass through all the branches. This is essentially a compositional reasoning task, in which the agent must aggregate information from each branch, and compose them into a set of unifying input conditions. For illustrative purposes, here are some of the conditions generated by the agent for our motivating example from \Cref{fig:motivating} and \Cref{sec:motivation}:
\begin{figure}[h!]
\begin{lstlisting}[basicstyle=\ttfamily, breaklines=true, breakindent=16pt, escapeinside={(*@}{@*)}]
1. The input must not be null...
2. The input must not be an empty string after trimming whitespace...
3. The input must not contain || ...
\end{lstlisting}
\end{figure}

\noindent
Since these conditions succinctly summarize the detailed branch information produced earlier, we collect these conditions for use in the next step of the tool and \textit{discard} the detailed branch information.

\subsection{Test Generation and Repair}
\label{sec:testgen}

At this stage, we have a detailed description of a flow from source to sink, along with a series of constraints that a test input must satisfy in order to exercise this flow. The final stage of our system involves using the flow information and input constraints to generate an initial proof-of-vulnerability test case, followed by feedback-driven repair.

\noindent
\textbf{Criteria for Success } To construct a framework that can generate a PoV test, we first have to understand what it means for a PoV test to be successful. This is surprisingly non-trivial. Consider our motivating example again (\Cref{fig:motivating}). The root cause of the vulnerability is that the error message can, in certain cases, reproduce the user-provided \TT{value} string verbatim. A hypothetical PoV test could call \TT{isValid} with a \TT{value} string embedded with a specific sequence of characters, say \TT{"abcd"}, in the appropriate format to reach the sink. It could then assert that the error message does not contain \TT{"abcd"}. Prima-facie, this would satisfy the requirements of a PoV test:
\be
\item It would \textit{fail} if and only if the vulnerability \textit{exists},
\item It would execute the vulnerable method and use the observed result directly in an assertion to check the presence of a bug.
\ee
However, this is a \textit{Code Injection} vulnerability, and the would-be PoV test \textit{does not actually inject any code}! This shows us that the definition of a successful PoV test must necessarily be tailored to the type of vulnerability, \eg a Code Injection vulnerability must inject code and execute it in addition to satisfying the above three conditions.

Previous works on exploit generation \cite{abramovichenigma, simsek2025pocgen} have attempted to group exploits based on their CWE category and develop criteria for each category that can be automatically checked. For example, an exploit for OS Command Injection (CWE-78) could be verified by checking whether the test can execute a specified binary like \TT{/usr/bin/mybin}. These are akin to Capture-The-Flag (CTF) challenges. However, these definitions can be too restrictive in our setting. For example, \TT{CVE-2014-3576} is an OS Command Injection vulnerability that allows an attacker to specifically execute the \TT{shutdown} command. A definition based on \TT{/usr/bin/mybin} would be a mismatch for this exploit.

We instead opt to use semantic criteria that can be manually checked. For each of the CWE categories we consider in this paper, we define what it means for an exploit of that category to be successful:
\bi
\item \textit{CWE-22 (Path Traversal):} The test case must use a public API of the project to read from or write to at least one file outside the project directory.
\item \textit{CWE-78 (OS Command Injection):} The test case must use a public API of the project to execute any shell command that is not intended by the application.
\item \textit{CWE-79 (Cross-Site Scripting):} The test case must call a public API of the project with an input that contains embedded scripting code, and show that this input is not sanitized properly.
\item \textit{CWE-94 (Code Injection):} The test case must call a public API of the project with an input that contains embedded code, and this code must be executed.
\ei
\res{I think we need to define these vulnerabilities before saying the success criteria}

\noindent
\textbf{Generating an initial test:} When humans write code to solve a task, we rarely write correct functional code in a single attempt. Instead, we usually follow an iterative process, where we write some code, observe its behavior, add debugging statements if necessary, re-run the code, and so on. We used this process flow as a guide while developing our test generation component. In addition to the tools mentioned in previous stages, \ie \texttt{ListDir}, \texttt{Read}, \texttt{Find} and \texttt{Grep}, we also give the agent the ability to write to files (\texttt{Write}) and run the current test code to observe its output (\texttt{Run}).

To ensure consistent environment-independent execution, we ask the agent to wrap each project and its dependencies as a Docker container. The container, when built and run, must execute the test case. The \texttt{Run} tool does not permit the execution of arbitrary terminal commands; rather, it just builds and runs the Docker container, and furnishes the agent with the output of these commands.

\Cref{fig:test_prompt} shows a portion of our test generation prompt for a CWE-94 (Code Injection) vulnerability. We include the flow information and input constraints obtained in \Cref{sec:dataflow} and \Cref{sec:controlflow} respectively. The prompt is specific to the CWE category of the vulnerability and includes the criteria for an exploit to be considered successful. We also add instructions in the prompt to avert certain common failure modes --- 1) tests which don't actually run the program, and instead match shallow patterns in the source code to check for the presence of certain text, 2) tests that simulate the vulnerability by re-implementing a simplified version of it, without running the existing project code. We explicitly instruct the agent to avoid these patterns of behavior.

Once the agent has generated a test and is satisfied that it runs correctly, we instruct it to respond \TT{<DONE>} to trigger the next phase.

\begin{figure}
\begin{lstlisting}[basicstyle=\small\ttfamily\color{dkgreen}, breaklines=true, breakindent=0pt, escapeinside={(*@}{@*)}, columns=fullflexible]
Create a test case that FAILS (exits with non-zero code) if the vulnerability EXISTS, and PASSES (exits with code 0) if the vulnerability DOES NOT EXIST.
(*@\hl{This is a Code Injection vulnerability (CWE-94). The test case must call a public API of the project with an input that contains embedded code, and this code must be executed.}@*)
This test should actually run the vulnerable code in the project.
- It should NOT read the source code to check for the presence of a vulnerability.
- It should NOT "simulate" the vulnerability by running some separate code that does not use the project.
Here is a flow consisting of a sequence of program points to reach the vulnerability:
(*@\textcolor{black}{\{flow\}}@*)
The test should start from the flow 'source' and reach the 'sink'. It should be designed such that it passes through all the branch conditions on the way. This means that the input and method calls should be carefully crafted, satisfying the following conditions:
(*@\textcolor{black}{\{input\_conditions\}}@*)
\end{lstlisting}
\caption{A portion of our test generation prompt for a CWE-94 (Code Injection) vulnerability. The highlighted portion is modified depending on the CWE category. For the full prompt, refer to the appendix.}
\label{fig:test_prompt}
\end{figure}

\noindent
\textbf{Feedback-driven repair:} We validate the generated test case with two automated checks. We first build the project as a Docker image, and check if it completes successfully. If it does, we run a container with the built image, and check that the run fails (exits with non-zero code). If either of these stages does not complete as expected, \ie if the build fails or the run succeeds, we collect the output from that stage and use it as feedback for the agent.

We prompt the agent with this feedback, instructing it to fix the test by carefully analyzing errors or messages in the output, and reasoning step by step about what might have gone wrong. We perform this feedback-driven repair in a loop until a preset maximum number of iterations is reached.

%% file: sections/4_setup.tex
\section{Experimental Setup}
\label{sec:setup}

\subsection{Benchmarks}
\label{sec:benchmarks}
CWE-Bench-Java \cite{li2025iris} is a dataset of 120 Java programs with known vulnerabilities. The vulnerabilities span 4 CWE categories --- Path Traversal, OS Command Injection, Cross-Site Scripting (XSS) and Code Injection. Each vulnerability includes metadata like its vulnerability report from the National Vulnerability Database\cite{nist_nvd_2024}, the URL of the GitHub repo of the project, buggy and fixed commit hashes, build instructions, and the classes and methods changed to fix the vulnerability. We attempted to clone each project and build it in both the vulnerable and fixed states, using the build scripts provided with the CWE-Bench-Java dataset. We were unable to fetch the fix commit or build at the fixed commit for some of these projects, leaving us with a filtered dataset of 70 Java programs.

PrimeVul \cite{ding2024vulnerability} is a dataset of over 7000 vulnerabilities in C and C++ programs. We filter these to include only vulnerabilities belonging to our selected CVE types, and select 30 vulnerabilities at random from the filtered set. However, PrimeVul does not annotate each project with build information. So this is possibly a more challenging setting for PoV test generation, where the model has to build the project successfully as a prerequisite for generating a test.

Our entire evaluation dataset therefore consists of 100 problem instances (70 from CWE-Bench-Java and 30 from PrimeVul) covering 3 programming languages (Java, C and C++).

\begin{table}[t]
    \begin{tabular}{l|c|c|c|}
    \clineB{2-4}{2}
    & \textbf{CWE-Bench-Java} & \textbf{PrimeVul} & \textbf{Total} 
    \bigstrut\\ \clineB{2-4}{2}
\multicolumn{1}{l}{} &   \multicolumn{1}{l}{}     & \multicolumn{1}{l}{}     & \multicolumn{1}{l}{}         \bigstrut\\[-1.2em]
    \clineB{1-4}{1}
    \multicolumn{1}{|l|}{Path Traversal} & 35 & 13 & 48 \bigstrut\\[0.25em] \clineB{1-4}{1}
    \multicolumn{1}{|l|}{Command Injection} & 6 & 4 & 16 \bigstrut\\[0.25em] \clineB{1-4}{1}
    \multicolumn{1}{|l|}{Cross-Site Scripting} & 15 & 3 & 18 \bigstrut\\[0.25em] \clineB{1-4}{1}
    \multicolumn{1}{|l|}{Code Injection} & 14 & 10 & 24 \bigstrut\\[0.25em] 
    \clineB{1-4}{2}
 \hlineB{2}
\multicolumn{1}{l}{} &   \multicolumn{1}{l}{}     & \multicolumn{1}{l}{}     & \multicolumn{1}{l}{}     \bigstrut\\[-1.2em] \hlineB{2}
\multicolumn{1}{|l|}{\textbf{Total}} & \textbf{70} & \textbf{30} & \textbf{100} \bigstrut\\\clineB{1-4}{2}
    \end{tabular}
    \vspace{5pt}
    \caption{The statistics of our benchmark datasets, showing the number of instances of each vulnerability type.}
    \label{tab:datasets}
\end{table}

\begin{figure*}[t]
    \centering
    \begin{subfigure}[t]{0.48\textwidth}
        \includegraphics[width=\linewidth]{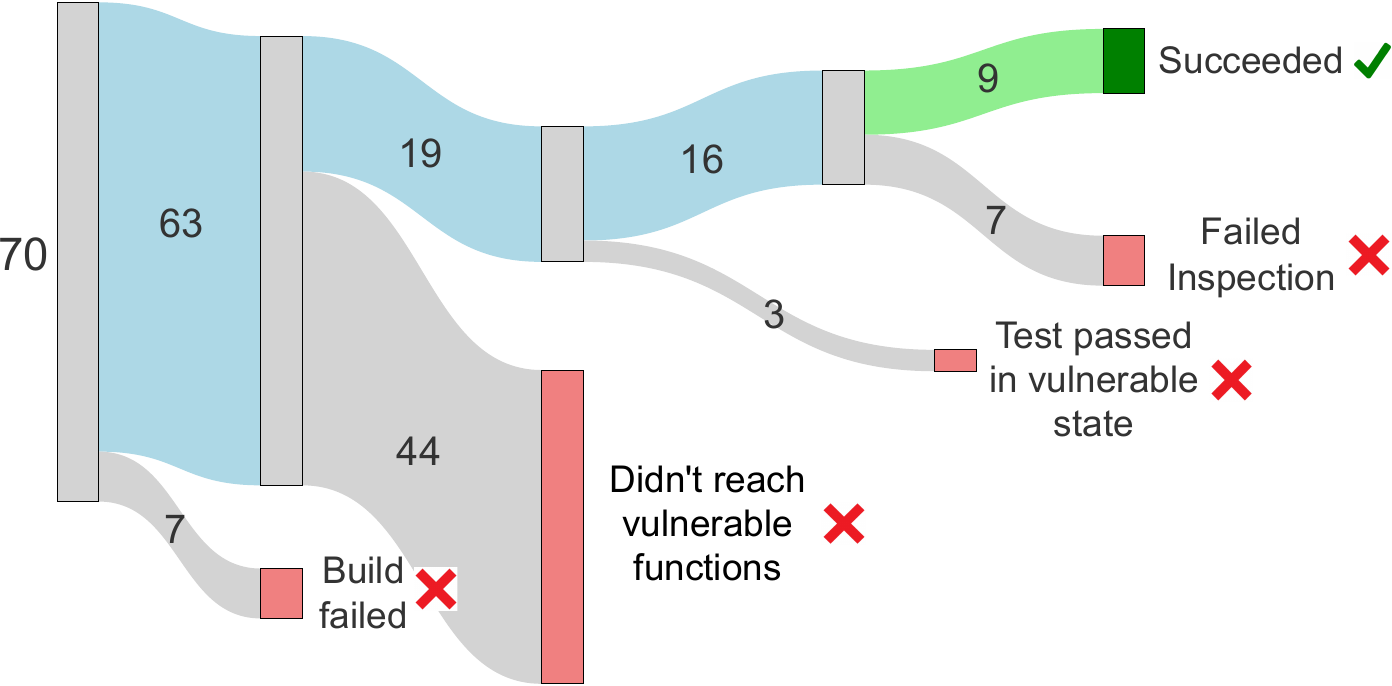}
        \caption{CodeAct 2.1, CWE-Bench-Java}
    \end{subfigure}
    \hfill
    \begin{subfigure}[t]{0.48\textwidth}
        \includegraphics[width=\linewidth]{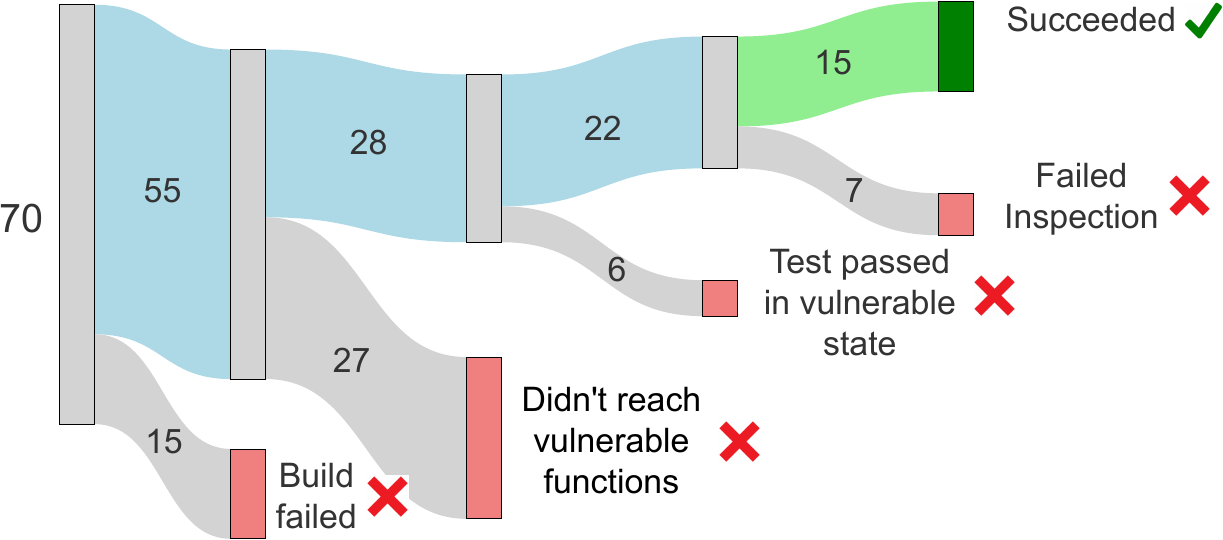}
        \caption{\tool{}, CWE-Bench-Java}
    \end{subfigure}
    ~\\
    ~\\
    ~\\
    \begin{subfigure}[t]{0.48\textwidth}
        \includegraphics[width=\linewidth]{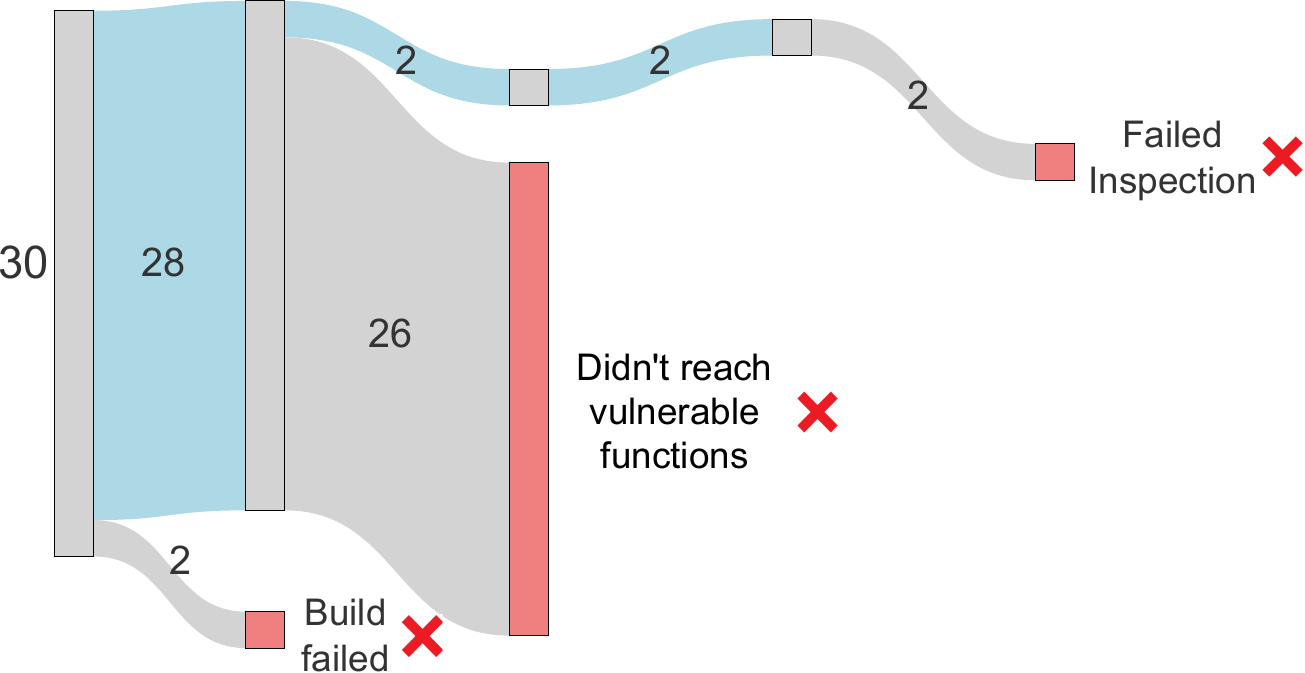}
        \caption{CodeAct 2.1, PrimeVul}
    \end{subfigure}
    \hfill
    \begin{subfigure}[t]{0.48\textwidth}
        \includegraphics[width=\linewidth]{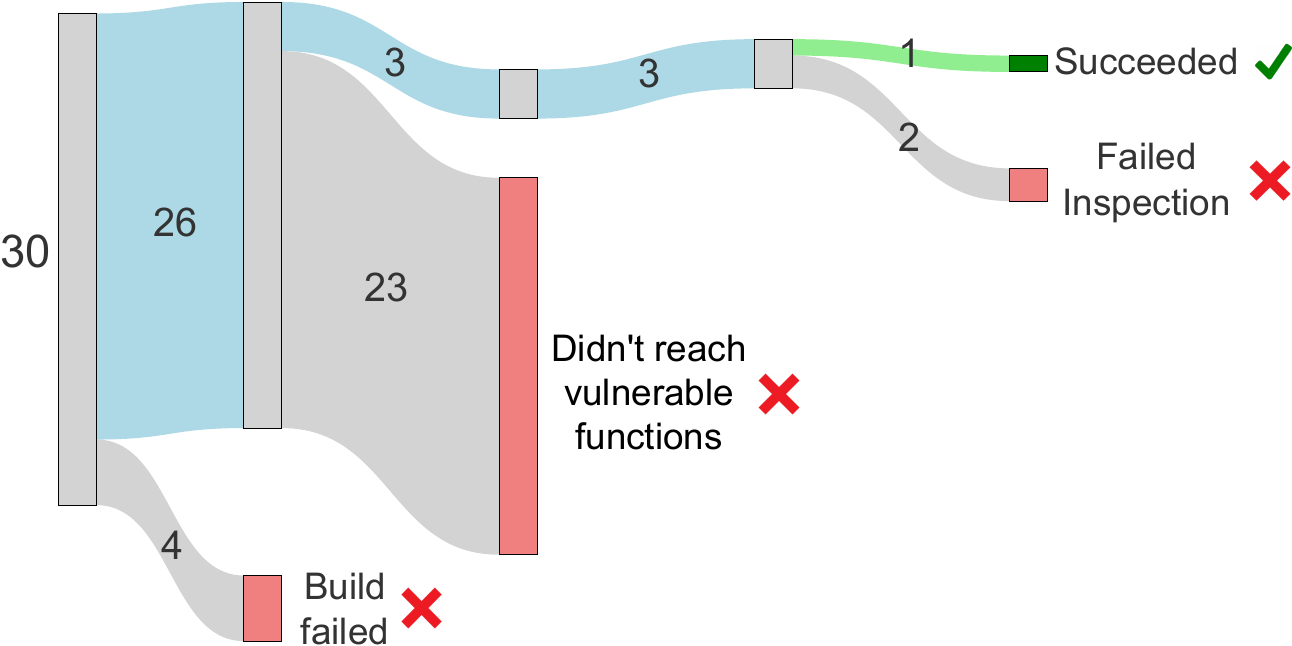}
        \caption{\tool{}, PrimeVul}
    \end{subfigure}    
    \vspace{8pt}
    \caption{A detailed analysis of the test generation performance of CodeAct 2.1 (on the left) vs \tool{} (on the right), for our two benchmarks CWE-Bench-Java and PrimeVul. \tool{} is able to generate more successful tests (15 vs 9 and 1 vs 0), as well as more tests that reach the vulnerable functions (28 vs 19 and 3 vs 2).}
    \label{fig:main_result}
\end{figure*}

\subsection{Implementation}
\label{sec:implementation}
\noindent
\textbf{Setting up benchmarks:} We reset each project to the vulnerable commit and build it as a Docker container. We make sure that the Dockerfile contains all the dependencies needed for the project, and does not require any external volumes. This ensures that a) the tests are run in a sandboxed environment, and b) the runs are reliably reproducible.

\noindent
\textbf{Implementing the system:} \tool{} is implemented in Python and run with Docker. We developed custom interfaces for all the tools we discussed in \Cref{sec:methodology}. The underlying LLM for all our agent calls is Claude-3.7-Sonnet, which we access through \TT{litellm}. We set a maximum budget of 5 USD and a time budget of 40 mins for each project.

\subsection{Baselines}
Our main baseline is the CodeAct 2.1 agent \cite{wang2024executable} running in the OpenHands framework \cite{wang2025openhands}. This is a general-purpose, open-source, software agent. Although there are many other LLM agents that are specialized for the task of fixing software bugs, fixing bugs is a \textit{very different problem} from generating tests. CodeAct 2.1 with OpenHands performs competitively with other specialized test generation models \cite{neubig2024-openhands-codeact-2.1:} on the SWE-Bench benchmark \cite{jimenez2023swe}, and its capabilities extend to any general software engineering task. It outperforms popular agents like SWE-Agent by a large margin \cite{neubig2024-opendevin-codeact-1.0:}. So it is a natural choice to use as a benchmark. Just as in \tool{}, we use Claude 3.7 Sonnet as the underlying LLM and impose the same budget and time constraints. The full prompt that we used for OpenHands is available in the appendix.

\subsection{Metrics}
\label{sec:metrics}

We follow a set of steps to evaluate the correctness of a PoV test. If any of these steps fails, we abort the evaluation and return failure. In order:
\be
\item \textbf{Build:} Build the project along with its created tests at the vulnerable commit, as a Docker image.
\item \textbf{Run:} Run the Docker image as a container, and check that it exits with non-zero code.
\item \textbf{Check coverage:} Check that the program flow in the previous step reached a method corresponding to the vulnerability. To evaluate this, we instrument each function to print its name when it is called. If the output contains the name of any method that was changed as part of the vulnerability fix, then we consider this satisfactory. This helps weed out tests that use shallow pattern matching against the text of the source program.
\ee
If a test passes all these 3 criteria, then the final step is to \textbf{manually inspect} it to evaluate whether it satisfies the category-specific criteria listed in \Cref{sec:testgen}. For example, if the vulnerability is Code Injection, then we verify that the test calls a public API of the project with an input that contains embedded code, and that the code is actually executed. If the test passes this check, we consider it correct.

% Sometimes, a test fails to build when the project is in the fixed state, because of mismatched dependencies or versions. This is not the agent's fault, because it does not have access to the fixed state of the project. For such cases, we manually update the test and re-run it. There are also certain cases where the test exits with a non-zero code in the fixed version because the project detects the attempted exploit and raises an exception. The test should ideally catch this exception and return success, but once again, the agent does not have access to the source code of the fixed version of the project. So it cannot know, a priori, the exact exception that will be thrown, or even the fact that an exception will be thrown at all. Therefore, we do not penalize the agent in such cases and consider the test correct.

%% file: sections/5_results.tex
\section{Experimental Results}
We evaluate our approach through the following research questions:
\bi
\item \textbf{RQ1: Performance of our tool.} How many PoV tests is \tool{} successfully able to generate? How does this compare with our baselines?
\item \textbf{RQ2: Different vulnerability types.} How does the performance of \tool{} vary across vulnerability types as represented by CWE categories?
\item \textbf{RQ3: Evaluating our design choices.} What is the impact of the flow reasoning and branch reasoning components on the performance of \tool{}?
% \item \textbf{RQ3: Accuracy of LLM-based program analysis.} We use an LLM to perform flow reasoning and branch reasoning. How accurate is the output of the LLM on these tasks?
\ei

\subsection{RQ1: Performance Evaluation}
\label{sec:rq1}
In this section, we measure the effectiveness of \tool{} in generating PoV tests, and measure its performance relative to our baselines.

\noindent
\textit{\textbf{Evaluation:}} We run \tool{} on our benchmark dataset of 100 programs. We also run CodeAct 2.1 in the same setting, with identical time and budget constraints. For each generated test case, we evaluate it according to the process defined in \Cref{sec:metrics}. If a test fails at a particular stage of this checking process, we collect that information.

\noindent
\textit{\textbf{Discussion:}} The overall results are in \Cref{fig:main_result}. We make the following observations:
\bi
\item \textit{CWE-Bench-Java results:} \tool{} is able to generate successful PoV tests for 15 out of 70 CWE-Bench-Java projects (\textbf{21\%}) compared to CodeAct which succeeds for only 9 (\textbf{13\%}). This is a significant gap, and clearly shows the benefits of our multi-stage agentic workflow.
\item \textit{PrimeVul results:} On the PrimeVul dataset, CodeAct is unable to solve any of the 30 problems, whereas \tool{} is able to solve 1. As mentioned in \Cref{sec:setup}, PrimeVul does not have build scripts for individual projects; however this (perhaps surprisingly) does not seem to pose a challenge to either CodeAct or \tool{}. Out of 30 projects, the resulting Dockerfiles build successfully for 27 and 26 of the projects, for CodeAct and \tool{} respectively. However, the test flow reaches the vulnerability for only 3 and 2 projects respectively. This suggests that test generation is particularly challenging on PrimeVul not because of a lack of build information, but because of a lack of understanding of the flow of each vulnerability. Further research is needed to develop better PoV test generation tools for these projects.
\item \textit{Vulnerable function coverage:} \tool{} is able to generate tests that reach vulnerable functions for \textbf{28} CWE-Bench-Java projects versus only \textbf{17} for CodeAct, and likewise for PrimeVul (\textbf{3} vs \textbf{2}). This is evidence that our flow and branch reasoning steps are having the desired effect, enhancing the ability of the agent to produce test inputs that reach vulnerability sinks.
\ei

\result{\tool{} is able to generate PoV tests for \textbf{16} vulnerabilities, as compared to just \textbf{9} for the CodeAct 2.1 baseline. Further, \tool{}-generated tests reach the vulnerable functions in \textbf{31} projects, as compared to \textbf{19} for the baseline.}

\begin{figure}
    \centering
    \includegraphics[width=0.9\linewidth]{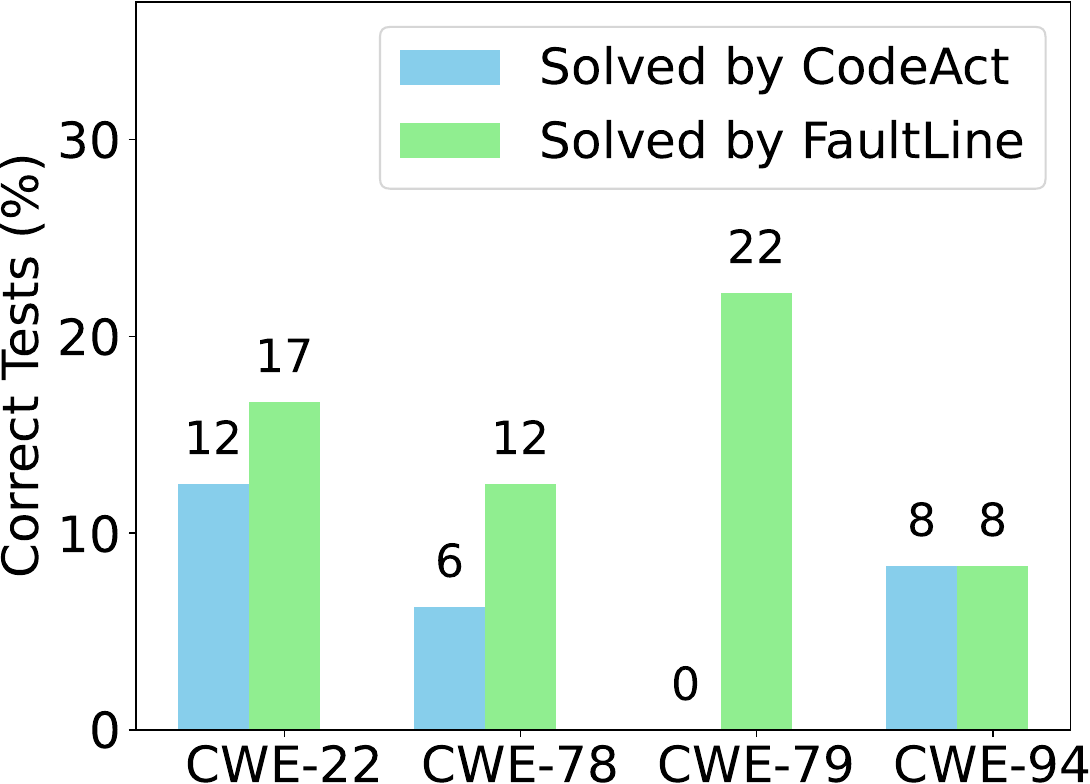}
    \vspace{4pt}
    \caption{Percentage of correct PoV tests per CWE category.}
    \label{fig:cwe_categories}
\end{figure}

\subsection{RQ2: Vulnerability Types}
\label{sec:rq2}
In this section, we evaluate how the performance of our agent varies across vulnerability types.

\noindent
\textit{\textbf{Evaluation:}} We collect the results of PoV test generation as described in \Cref{sec:rq1}, and categorize each test according to the CWE-ID of the project it corresponds to. We plot the percentage of successful tests in each category, for both tools.

\noindent
\textit{\textbf{Discussion:}} The results are shown in \Cref{fig:cwe_categories}. We make the following observations:
\bi
\item As measured by average percentage of successful PoV tests generated, CWE-94 (Code Injection) is the hardest category, with both tools reporting a success rate of just 8\%. CWE-22 (Path Traversal) and CWE-79 (Cross-Site Scripting) seem to be relatively easier by comparison, but overall it is clear that this is a challenging task irrespective of category. 
\item Another observation is that for each category, \tool{} is consistently on par with or better than CodeAct 2.1. This shows that its performance gains are not specific to any one type of vulnerability. Although in this paper we have only evaluated our technique on 4 CWE categories, based on these results we expect it to generalize well to other categories too.
\ei

\result{\tool{} performs on par or improves over the baseline for all 4 CWE categories, showing its generalizability.}

\subsection{RQ3: Evaluating Design Choices}
\label{sec:rq3}
In this section, we evaluate the impact of the flow reasoning and branch reasoning components on the performance of \tool{}.

\begin{table}[t]
    \begin{tabular}{lc|}
    \clineB{2-2}{2}
    & \multicolumn{1}{|c|}{\textbf{Successful PoV Tests}}
    \bigstrut\\ \clineB{2-2}{2}
     & \multicolumn{1}{l}{}      \bigstrut\\[-1.2em]
    \clineB{1-2}{1}
    \multicolumn{1}{|l|}{w/o Flow, w/o Branch} & 11 \bigstrut\\[0.25em] \clineB{1-2}{1}
    \multicolumn{1}{|l|}{w/o Flow} & 9 \bigstrut\\[0.25em] \clineB{1-2}{1}
    \multicolumn{1}{|l|}{w/o Branch} & 11 \bigstrut\\[0.25em] \clineB{1-2}{2}
 \hlineB{2}
     & \multicolumn{1}{l}{}        \bigstrut\\[-1.2em] \hlineB{2}
\multicolumn{1}{|l|}{\tool{}} & \textbf{16} \bigstrut\\\clineB{1-2}{2}
    \end{tabular}
    \vspace{5pt}
    \caption{Assessing the impact of the flow and branch reasoning steps.}
    \label{tab:ablation}
\end{table}

\noindent
\textit{\textbf{Evaluation:}} We design three configurations of the tool by removing flow analysis, branch analysis, and both flow and branch analysis. Running the each of these configurations on all 100 examples was prohibitively expensive, so instead, we selected the 16 projects corresponding to tests generated by \tool{} that passed evaluation, and ran each tool configuration on this subset of 16 projects.

\noindent
\textit{\textbf{Discussion:}} The results are shown in \Cref{tab:ablation}. We can see that \tool{} is the best performing configuration, and that the configurations without flow and branch analysis do not generate as many successful tests. This shows that each component of our system is crucial to its overall performance.
One interesting observation is that branch reasoning alone, without flow reasoning, performs worse than a configuration with neither branch nor flow reasoning. In this configuration, we are essentially prompting the model to extract branch conditions along a unspecified data flow, and further, reason about conditions that an unspecified source input has to satisfy. This requires two reasoning steps, and intuitively, scaffolding the reasoning process should yield better results.

To further investigate the effectiveness of our multi-step reasoning pipeline, we inspect the system's outputs for our motivating example from \Cref{sec:motivation}. The system generates the correct flow from source to sink, shown in the top half of \Cref{fig:output_example}.
\begin{figure}
\begin{lstlisting}[basicstyle=\small\ttfamily\color{dkgreen}, breaklines=true, breakindent=0pt, escapeinside={(*@}{@*)}, columns=fullflexible]
<FLOW>
{"role": "Source",
 "code": "public boolean isValid(String (*@\hl{value}@*), ConstraintValidatorContext context) {",
 "variable": "value",
 "file": "...CronValidator.java",
 "remarks": "The entry point where untrusted cron expressions are received for validation"
}
...
{"role": "Sink",
 "code": "context.buildConstraintViolationWithTemplate((*@\hl{e.getMessage()}@*)).addConstraintViolation();",
 "variable": "e.getMessage()",
 "file": "...CronValidator.java",
 "remarks": "The unvalidated input is used in a template context, allowing for Java EL expression injection"
}
</FLOW>
<CONDITIONS>
...
6. The input must have a (*@\hl{valid number of parts}@*) that matches one of the expected cron expression formats
7. The input must contain at least one special character (/, -, or ,) to trigger the complex parsing path
...
<CONDITIONS>
\end{lstlisting}
\caption{A portion of the output of flow and branch reasoning of \tool{} for the motivating example from \Cref{sec:motivation}. Deriving the right conditions on the input string allows the test-generation agent to generate a PoV test that reaches the vulnerability sink.}
\label{fig:output_example}
\end{figure}
Then, using this flow, \tool{} extracts branch conditions and generates input constraints, a portion of which are shown in the bottom half of \Cref{fig:output_example}.

Notice that this contains the constraint ``the input must have a valid number of parts''. As we discussed in \Cref{sec:motivation}, CodeAct fails to understand this requirement and generates an input containing just the wrong number of space-separated components. On the other hand, \tool{} is able to use these constraints to generate a test that calls \TT{isValid} with the following string: \TT{"* * * * * \$\{T(java.lang.Runtime).getRuntime().exec('touch /tmp/abc"')\}"}. This has 7 space-separated components, which is a valid format and allows the dangerous input to reach the vulnerability sink without sanitization.

\result{Without its flow or branch reasoning components, \tool{} solves between 9 and 11 problems, as compared to 16 in its full configuration. This shows the utility of each component.}

% \subsection{RQ4: LLM-Based Program Analysis}
% \label{sec:rq3}
% In our system, we use an LLM to perform flow and branch analysis, and we have shown that these steps have a positive impact on the overall performance. However, we would now like to investigate their accuracy, standalone.  

% \noindent
% \textit{\textbf{Evaluation:}}
% \red{In how many cases does the generated flow correspond to vulnerable functions?}

% \noindent
% \textit{\textbf{Discussion:}}

% \result{Summary goes here}

%% file: sections/6_related.tex
\section{Related Work}

\textbf{Vulnerability Datasets:}
There are many curated vulnerability datasets in different languages, some with proof-of-vulnerability tests. BigVul~\cite{fan2020ac} is a dataset of 3,754 C/C++ vulnerabilities, and PrimeVul~\cite{ding2024vulnerability} builds on BigVul to create a higher quality dataset of 6,968 C/C++ vulnerabilities. CrossVul~\cite{nikitopoulos2021crossvul} has $\sim$13,000 vulnerabilities in 40 programming languages. SVEN~\cite{he2023large} consists of $\sim$1600 C/C++/Python vulnerabilities. These datasets include, for each vulnerability, the URL of the GitHub repository from which the vulnerability was sourced, the commit message corresponding to the fix, and the patch. However, these datasets lack information on how to build each project and reproduce the vulnerability. They also lack proof-of-vulnerability test cases.

CWE-Bench-Java~\cite{li2025iris} is a dataset of 120 Java vulnerabilities with build information for each project, but no proof-of-vulnerability test cases. Vul4J~\cite{bui2022vul4j} is a small dataset of 79 Java vulnerabilities, each of which is reproducible and has proof-of-vulnerability test cases. ARVO~\cite{mei2024arvo} is a continually expanding database of C/C++ vulnerabilities collected from Google's OSS-Fuzz tool. SecBench.js~\cite{bhuiyan2023secbench} is a dataset of JavaScript vulnerabilities, some of which have exploit code.

\noindent
\textbf{LLM Agents for Bug Reproduction:}
There is a line of research on generating bug reproduction tests from reports. LIBRO~\cite{kang2024evaluating} frames this as a few-shot code generation problem, where an LLM is shown examples of bug reports and corresponding tests, before being asked to generate a test for a given report. \citet{cheng2025agentic} build on the LIBRO framework by designing an agentic system with a fine-tuned code editing tool. Otter~\cite{ahmed2024tdd,ahmed2025otter} is an LLM agent workflow that uses systematic reasoning to generate bug reproduction tests. However, this is qualitatively very different from vulnerability exploit generation, which involves generating carefully crafted inputs that can traverse long sequences of method calls within a program.

\noindent
\textbf{Proof-of-Vulnerability Test Generation:}
There is a long line of work that predates LLMs~\cite{brumley2008automatic,avgerinos2014automatic,hu2015automatic,alhuzali2016chainsaw,huang2013craxweb}, based on deriving constraints on a program's input and using symbolic execution to solve these constraints. However, these are specialized to certain kinds of vulnerabilities, and cannot reach vulnerabilities that are deeply embedded in a program. SemFuzz~\cite{you2017semfuzz} is a fuzzing tool that uses semantic information from vulnerability reports to perform guided fuzzing. However, this can only generate relatively simple kinds of inputs, and can only detect vulnerabilities that result in runtime errors like crashes, resource leaks or infinite loops. ARVO~\cite{mei2024arvo} collects bug reports from Google's OSS-Fuzz~\cite{serebryany2017oss}, derives build and dependency information for each project, and creates a dataset of reproducible vulnerabilities. However, ARVO relies on the bug report already containing the exact input that triggers the vulnerability.

There has been research on using LLM agents to generate exploits for web applications in a one-day~\cite{fang2024llm} and zero-day~\cite{zhu2024teams} setting. \citet{zhu2025cve} design a comprehensive benchmark for evaluating exploit generation in web applications. However, the setting of these works is significantly different from ours, because exploiting web vulnerabilities requires interacting with a webpage, \eg entering text in a box or clicking a button; as opposed to our setting which requires writing code that calls an API. EniGMA~\cite{abramovichenigma} augments an SWE-Agent~\cite{yang2024swe} with additional tools like an interactive debugger, to enable it to solve Capture-The-Flag (CTF) problems. The solution to a CTF problem is a string, or ``flag'', that has to be retrieved. This is qualitatively very different from our setting, which involves writing test cases.

PoCGen~\cite{simsek2025pocgen} is a concurrent work to ours that involves generating proof-of-concept exploits for vulnerabilities in NPM packages. However, it relies on static and dynamic analysis tools for NPM, which limits its generalizability across programming languages.

%% file: sections/7_threats.tex
\section{Limitations and Threats}
As with any experimental study, the conclusions of our work must be considered in the context of the following potential threats to validity.

\noindent
\textbf{Data leakage:} The knowledge cutoff date of Claude-3.7-Sonnet, our base LLM, is November 2024. Our data sources for vulnerabilities, CWE-Bench-Java and PrimeVul, were curated before this date. It is very likely that the LLM has seen several of these vulnerabilities as part of its training data. Therefore, it is possible that our careful prompting is not actually \textit{eliciting reasoning}, but simply enabling the model to \textit{recall instances} from its training data. This threat is somewhat mitigated by two factors:
\be
\item PoV tests are seldom made public for security-related reasons. Although the model may have seen several of the vulnerabilities in its training data, it is unlikely to have seen the corresponding PoV tests.
\item Our primary comparison is with the CodeAct agent, which uses the same underlying LLM. Therefore, there is no unfair advantage gained by our agentic framework compared to the baseline.
\ee

\noindent
\textbf{Test success in fixed state:} An ideal proof of vulnerability test must not only fail when the vulnerability exists, but also \textit{pass} when it is fixed. However, our evaluation criteria do not include a check that the test passes in the fixed state. When we tried running our tests on the fixed versions of each project, we observed certain issues:
\be
\item Sometimes, a test fails to build when the project is in the fixed state, because of mismatched dependencies or versions. This is not the agent's fault, because it does not have access to the fixed state of the project.
\item There are also certain cases where the test exits with a non-zero code in the fixed version because the project detects the attempted exploit and raises an exception. The test should ideally catch this exception and return success, but once again, the agent does not have access to the source code of the fixed version of the project. So it cannot know, a priori, the exact exception that will be thrown, or even the fact that an exception will be thrown at all. 
\ee
For all of these reasons, we choose not to include passing in the fixed state as a criterion for a successful PoV test, and we defer the question of how to accomplish this to future work.

\noindent
\textbf{Manual Inspection:} The final step of our evaluation procedure (\Cref{sec:metrics}) is a manual inspection. Although this is based on objectively verifiable criteria, there is a possibility of errors in human judgment during the labeling process. This is somewhat mitigated by the fact that the number of examples necessitating such manual labeling is low, and we made every effort to be thorough with our analysis, but nevertheless it remains a limitation of our experimental design.

\noindent
\textbf{Generalizability:} Finally, we acknowledge that \tool{} is evaluated on only 4 CWE categories, which means that our conclusions have to be contextualized accordingly. However, our system design does not make any assumptions on the type of vulnerability, and we see consistent gains over the baseline across all our 4 categories. Thus, we expect that the conclusions would hold for other categories too, but we defer this investigation to future work.

%% file: sections/8_conclusion.tex
\section{Conclusion}
Proof-of-vulnerability tests are of vital importance to enable developers to understand a vulnerability and avoid regressions. Generating these tests involves subtle reasoning about program properties, and proves extremely challenging for state-of-the-art LLM agents. In this paper, we have developed \tool{}, a system that utilizes carefully designed LLM reasoning steps to automatically generate PoV tests. Our results highlight the effectiveness of multi-step reasoning workflows in LLM agents, and our benchmark of 100 projects represents a challenging direction for further research in LLM agents and test generation.

%% file: sections/supplementary.tex
\section{Prompts}

\begin{lstlisting}[basicstyle=\small\ttfamily\color{black}, breaklines=true, breakindent=0pt, escapeinside={(*@}{@*)}, columns=fullflexible, caption=System Prompt]
You are a helpful AI assistant that can interact with a computer to solve tasks.

<ROLE>
Your primary role is to assist users by executing commands, modifying code, and solving technical problems effectively.
You should be thorough, methodical, and prioritize quality over speed.
Your code will never be read by humans, so focus on correctness, not style.
</ROLE>

<EFFICIENCY>
* Each action you take is somewhat expensive. Minimize unnecessary actions.
* When exploring the codebase, use the find and grep tools with appropriate filters to minimize unnecessary operations.
* You do not have access to the internet, so do not attempt to search online for information.
</EFFICIENCY>

<CODE_QUALITY>
* Write clean, efficient code with minimal comments. Avoid redundancy in comments: Do not repeat information that can be easily inferred from the code itself.
* When implementing solutions, focus on making the minimal changes needed to solve the problem.
* Before implementing any changes, first thoroughly understand the codebase through exploration.
* If you are adding a lot of code to a function or file, consider splitting the function or file into smaller pieces when appropriate.
</CODE_QUALITY>

<PROBLEM_SOLVING_WORKFLOW>
1. EXPLORATION: Thoroughly explore relevant files and understand the context before proposing solutions
2. ANALYSIS: Consider multiple approaches and select the most promising one
3. IMPLEMENTATION: Make focused, minimal changes to address the problem
</PROBLEM_SOLVING_WORKFLOW>

<TROUBLESHOOTING>
* If you've made repeated attempts to solve a problem but tests still fail or the user reports it's still broken:
  1. Step back and reflect on 5-7 different possible sources of the problem
  2. Assess the likelihood of each possible cause
  3. Methodically address the most likely causes, starting with the highest probability
  4. Document your reasoning process
</TROUBLESHOOTING>
\end{lstlisting}

\begin{lstlisting}[basicstyle=\small\ttfamily\color{black}, breaklines=true, breakindent=0pt, escapeinside={(*@}{@*)}, columns=fullflexible, caption=Flow Reasoning]
The project I am working with has a vulnerability, reported as a CWE. The issue description says:
{description}
You do not have access to the internet or GitHub to look up more details.
There are no vulnerability reports in the project directory either.

{tool_description}

Could you generate a sequence of program points to reach the vulnerable point (sink), starting from an external input (source)? This corresponds to a vulnerable "flow" through the program.
The flow should take the form of a sequence of program points, each in the following format:
{"role": "Source|Intermediate|Sink",
 "code": "Source code of program point (1-2 lines),
 "variable": "Variable name",
 "file": "File path (absolute)",
 "remarks": "Comments about this point, if any"}
 You can use multiple intermediate steps and tool invocations, but when you are finished,
 your final response should contain the flow in the above format, within the tags <FLOW> and </FLOW>.
\end{lstlisting}

\begin{lstlisting}[basicstyle=\small\ttfamily\color{black}, breaklines=true, breakindent=0pt, escapeinside={(*@}{@*)}, columns=fullflexible, caption=Branch Reasoning Part 1]
The project I am working with has a vulnerability, reported as a CWE. The issue description says:
{description}
You do not have access to the internet or GitHub to look up more details.
There are no vulnerability reports in the project directory either.

{tool_description}

Here is a flow consisting of a sequence of program points to reach the vulnerability:
{flow}

Could you generate the sequence of branch conditions
encountered on the way to the sink, starting from the source?
Include *every single* if-else, try-except, or switch statement that the program flow will encounter in the path from the source to the sink.
This should take the form of a sequence of program points, each in the following format:
{"type": "If-Else | Try-Except | Switch",
 "code": "Source code of program point (1-2 lines),
 "file": "File path (absolute)",
 "outcome": "What should be the outcome of the branch statement in order to reach the vulnerability?"}
You can use multiple intermediate steps and tool invocations, but when you are finished,
your final response should contain the sequence in the above format, within the tags <SEQUENCE> and </SEQUENCE>.
\end{lstlisting}

\begin{lstlisting}[basicstyle=\small\ttfamily\color{black}, breaklines=true, breakindent=0pt, escapeinside={(*@}{@*)}, columns=fullflexible, caption=Branch Reasoning Part 2]
<Conversation continued from Part 1>
Based on the above branch conditions that you generated, infer a set of conditions that the external input must satisfy in order to reach the vulnerability. Your final answer should be in the following format:
<CONDITIONS>
1. Condition 1
2. Condition 2
...
</CONDITIONS>
\end{lstlisting}

\begin{lstlisting}[basicstyle=\small\ttfamily\color{black}, breaklines=true, breakindent=0pt, escapeinside={(*@}{@*)}, columns=fullflexible, caption=Test Generation]
The project I am working with has a vulnerability, reported as a CWE. The issue description says:
{description}
You do not have access to the internet or GitHub to look up more details. There are no vulnerability reports in the project directory either.

Now create a test case that FAILS (exits with non-zero code) if the vulnerability EXISTS,
and PASSES (exits with code 0) if the vulnerability DOES NOT EXIST.
{cwe_desc}
This test should actually run the vulnerable code in the project.
- It should NOT read the source code to check for the presence of a vulnerability.
- It should NOT \"simulate\" the vulnerability by running some separate code that does not use the project.

Here is a flow consisting of a sequence of program points to reach the vulnerability:
{flow}

The test should start from the vulnerability 'source' and reach the 'sink'. It should be designed such that it passes through all the branch conditions on the way. This means that the input and method calls should be carefully crafted, satisfying the following conditions:
{conditions}

The project is built and run as a Docker container, and the Dockerfile is at `{workdir}/Dockerfile.vuln`. All the build dependencies for the project are already installed in `Dockerfile.vuln`. However, if you need any new dependencies, you can add them to `Dockerfile.vuln`.
Make sure to not modify any of the lines in the Dockerfile above \"# Do not modify anything above this line\". The entire project directory is copied into the Docker container, so you don't need to write any new COPY commands in the Dockerfile. The command to run the test should be the `CMD` command in `Dockerfile.vuln`, so that the test can be run with
`docker run -t imagename`.

Feel free to create any new files to create the test case.
You are highly encouraged to insert print statements in the existing source files to debug your test.
Remember the branch conditions and flow that you derived earlier, and use them to guide your test generation and debugging process.

Once you verify that the flow has reached the 'sink', you should analyze the observed behavior of the program to ensure that the test FAILS if the vulnerability exists, and PASSES if it does not exist. To re-emphasize, this test should NOT be based on reading the source code, but rather on the actual behavior of the program when it is run.
If I fix the vulnerability in the project, the test should PASS.

{tool_description}

If you successfully generate the test case and confirm that it satisfies all the above conditions, respond <DONE>.
\end{lstlisting}

\begin{lstlisting}[basicstyle=\small\ttfamily\color{black}, breaklines=true, breakindent=0pt, escapeinside={(*@}{@*)}, columns=fullflexible, caption=Repair]
The test you generated had the following error:
{feedback}
Please fix the test case. Carefully analyze this output for errors or messages that can help you debug your test. Reason step-by-step about what might have gone wrong, and how you can fix it.
You can use the <TOOL>...</TOOL> format to invoke tools, and you can also add new files.
When you have generated, run and checked your test again, respond with a message containing the string "<DONE>".
Remember that the test should actually run the vulnerable code in the project,
- It should NOT read the source code to check for the presence of a vulnerability.
- It should NOT \"simulate\" the vulnerability by running some separate code that does not use the project.
\end{lstlisting}

%% file: main.bbl

\begin{thebibliography}{36}

%%% the Web address.

\ifx \showCODEN    \undefined \def \showCODEN     #1{\unskip}     \fi
\ifx \showISBNx    \undefined \def \showISBNx     #1{\unskip}     \fi
\ifx \showISBNxiii \undefined \def \showISBNxiii  #1{\unskip}     \fi
\ifx \showISSN     \undefined \def \showISSN      #1{\unskip}     \fi
\ifx \showLCCN     \undefined \def \showLCCN      #1{\unskip}     \fi
\ifx \shownote     \undefined \def \shownote      #1{#1}          \fi
\ifx \showarticletitle \undefined \def \showarticletitle #1{#1}   \fi
\ifx \showURL      \undefined \def \showURL       {\relax}        \fi
% invisible to TeX
\providecommand\bibfield[2]{#2}
\providecommand\bibinfo[2]{#2}
\providecommand\natexlab[1]{#1}
\providecommand\showeprint[2][]{arXiv:#2}

\bibitem[cod({[n.\,d.]})]%
        {codeql}
 \bibinfo{year}{[n.\,d.]}\natexlab{}.
\newblock
\urldef\tempurl%
\url{https://codeql.github.com/}
\showURL{%
\tempurl}


\bibitem[Abramovich et~al\mbox{.}({[n.\,d.]})]%
        {abramovichenigma}
\bibfield{author}{\bibinfo{person}{Talor Abramovich}, \bibinfo{person}{Meet Udeshi}, \bibinfo{person}{Minghao Shao}, \bibinfo{person}{Kilian Lieret}, \bibinfo{person}{Haoran Xi}, \bibinfo{person}{Kimberly Milner}, \bibinfo{person}{Sofija Jancheska}, \bibinfo{person}{John Yang}, \bibinfo{person}{Carlos~E Jimenez}, \bibinfo{person}{Farshad Khorrami}, {et~al\mbox{.}}} \bibinfo{year}{[n.\,d.]}\natexlab{}.
\newblock \showarticletitle{EnIGMA: Interactive Tools Substantially Assist LM Agents in Finding Security Vulnerabilities}. In \bibinfo{booktitle}{\emph{Forty-second International Conference on Machine Learning}}.
\newblock


\bibitem[Ahmed et~al\mbox{.}(2025)]%
        {ahmed2025otter}
\bibfield{author}{\bibinfo{person}{Toufique Ahmed}, \bibinfo{person}{Jatin Ganhotra}, \bibinfo{person}{Rangeet Pan}, \bibinfo{person}{Avraham Shinnar}, \bibinfo{person}{Saurabh Sinha}, {and} \bibinfo{person}{Martin Hirzel}.} \bibinfo{year}{2025}\natexlab{}.
\newblock \showarticletitle{Otter: Generating Tests from Issues to Validate SWE Patches}.
\newblock \bibinfo{journal}{\emph{arXiv preprint arXiv:2502.05368}} (\bibinfo{year}{2025}).
\newblock


\bibitem[Ahmed et~al\mbox{.}(2024)]%
        {ahmed2024tdd}
\bibfield{author}{\bibinfo{person}{Toufique Ahmed}, \bibinfo{person}{Martin Hirzel}, \bibinfo{person}{Rangeet Pan}, \bibinfo{person}{Avraham Shinnar}, {and} \bibinfo{person}{Saurabh Sinha}.} \bibinfo{year}{2024}\natexlab{}.
\newblock \showarticletitle{TDD-Bench Verified: Can LLMs Generate Tests for Issues Before They Get Resolved?}
\newblock \bibinfo{journal}{\emph{arXiv preprint arXiv:2412.02883}} (\bibinfo{year}{2024}).
\newblock


\bibitem[Alhuzali et~al\mbox{.}(2016)]%
        {alhuzali2016chainsaw}
\bibfield{author}{\bibinfo{person}{Abeer Alhuzali}, \bibinfo{person}{Birhanu Eshete}, \bibinfo{person}{Rigel Gjomemo}, {and} \bibinfo{person}{VN Venkatakrishnan}.} \bibinfo{year}{2016}\natexlab{}.
\newblock \showarticletitle{Chainsaw: Chained automated workflow-based exploit generation}. In \bibinfo{booktitle}{\emph{Proceedings of the 2016 ACM SIGSAC Conference on Computer and Communications Security}}. \bibinfo{pages}{641--652}.
\newblock


\bibitem[Avgerinos et~al\mbox{.}(2014)]%
        {avgerinos2014automatic}
\bibfield{author}{\bibinfo{person}{Thanassis Avgerinos}, \bibinfo{person}{Sang~Kil Cha}, \bibinfo{person}{Alexandre Rebert}, \bibinfo{person}{Edward~J Schwartz}, \bibinfo{person}{Maverick Woo}, {and} \bibinfo{person}{David Brumley}.} \bibinfo{year}{2014}\natexlab{}.
\newblock \showarticletitle{Automatic exploit generation}.
\newblock \bibinfo{journal}{\emph{Commun. ACM}} \bibinfo{volume}{57}, \bibinfo{number}{2} (\bibinfo{year}{2014}), \bibinfo{pages}{74--84}.
\newblock


\bibitem[Bhuiyan et~al\mbox{.}(2023)]%
        {bhuiyan2023secbench}
\bibfield{author}{\bibinfo{person}{Masudul Hasan~Masud Bhuiyan}, \bibinfo{person}{Adithya~Srinivas Parthasarathy}, \bibinfo{person}{Nikos Vasilakis}, \bibinfo{person}{Michael Pradel}, {and} \bibinfo{person}{Cristian-Alexandru Staicu}.} \bibinfo{year}{2023}\natexlab{}.
\newblock \showarticletitle{SecBench. js: An executable security benchmark suite for server-side JavaScript}. In \bibinfo{booktitle}{\emph{2023 IEEE/ACM 45th International Conference on Software Engineering (ICSE)}}. IEEE, \bibinfo{pages}{1059--1070}.
\newblock


\bibitem[Brumley et~al\mbox{.}(2008)]%
        {brumley2008automatic}
\bibfield{author}{\bibinfo{person}{David Brumley}, \bibinfo{person}{Pongsin Poosankam}, \bibinfo{person}{Dawn Song}, {and} \bibinfo{person}{Jiang Zheng}.} \bibinfo{year}{2008}\natexlab{}.
\newblock \showarticletitle{Automatic patch-based exploit generation is possible: Techniques and implications}. In \bibinfo{booktitle}{\emph{2008 IEEE Symposium on Security and Privacy (sp 2008)}}. IEEE, \bibinfo{pages}{143--157}.
\newblock


\bibitem[Bui et~al\mbox{.}(2022)]%
        {bui2022vul4j}
\bibfield{author}{\bibinfo{person}{Quang-Cuong Bui}, \bibinfo{person}{Riccardo Scandariato}, {and} \bibinfo{person}{Nicol{\'a}s E~D{\'\i}az Ferreyra}.} \bibinfo{year}{2022}\natexlab{}.
\newblock \showarticletitle{Vul4J: a dataset of reproducible Java vulnerabilities geared towards the study of program repair techniques}. In \bibinfo{booktitle}{\emph{Proceedings of the 19th International Conference on Mining Software Repositories}}. \bibinfo{pages}{464--468}.
\newblock


\bibitem[Cheng et~al\mbox{.}(2025)]%
        {cheng2025agentic}
\bibfield{author}{\bibinfo{person}{Runxiang Cheng}, \bibinfo{person}{Michele Tufano}, \bibinfo{person}{J{\"u}rgen Cito}, \bibinfo{person}{Jos{\'e} Cambronero}, \bibinfo{person}{Pat Rondon}, \bibinfo{person}{Renyao Wei}, \bibinfo{person}{Aaron Sun}, {and} \bibinfo{person}{Satish Chandra}.} \bibinfo{year}{2025}\natexlab{}.
\newblock \showarticletitle{Agentic Bug Reproduction for Effective Automated Program Repair at Google}.
\newblock \bibinfo{journal}{\emph{arXiv preprint arXiv:2502.01821}} (\bibinfo{year}{2025}).
\newblock


\bibitem[Ding et~al\mbox{.}(2024)]%
        {ding2024vulnerability}
\bibfield{author}{\bibinfo{person}{Yangruibo Ding}, \bibinfo{person}{Yanjun Fu}, \bibinfo{person}{Omniyyah Ibrahim}, \bibinfo{person}{Chawin Sitawarin}, \bibinfo{person}{Xinyun Chen}, \bibinfo{person}{Basel Alomair}, \bibinfo{person}{David Wagner}, \bibinfo{person}{Baishakhi Ray}, {and} \bibinfo{person}{Yizheng Chen}.} \bibinfo{year}{2024}\natexlab{}.
\newblock \showarticletitle{Vulnerability detection with code language models: How far are we?}
\newblock \bibinfo{journal}{\emph{arXiv preprint arXiv:2403.18624}} (\bibinfo{year}{2024}).
\newblock


\bibitem[Fan et~al\mbox{.}(2020)]%
        {fan2020ac}
\bibfield{author}{\bibinfo{person}{Jiahao Fan}, \bibinfo{person}{Yi Li}, \bibinfo{person}{Shaohua Wang}, {and} \bibinfo{person}{Tien~N Nguyen}.} \bibinfo{year}{2020}\natexlab{}.
\newblock \showarticletitle{AC/C++ code vulnerability dataset with code changes and CVE summaries}. In \bibinfo{booktitle}{\emph{Proceedings of the 17th international conference on mining software repositories}}. \bibinfo{pages}{508--512}.
\newblock


\bibitem[Fang et~al\mbox{.}(2024)]%
        {fang2024llm}
\bibfield{author}{\bibinfo{person}{Richard Fang}, \bibinfo{person}{Rohan Bindu}, \bibinfo{person}{Akul Gupta}, {and} \bibinfo{person}{Daniel Kang}.} \bibinfo{year}{2024}\natexlab{}.
\newblock \showarticletitle{Llm agents can autonomously exploit one-day vulnerabilities}.
\newblock \bibinfo{journal}{\emph{arXiv preprint arXiv:2404.08144}}  \bibinfo{volume}{13} (\bibinfo{year}{2024}), \bibinfo{pages}{14}.
\newblock


\bibitem[Harzevili et~al\mbox{.}(2023)]%
        {harzevili2023survey}
\bibfield{author}{\bibinfo{person}{Nima~Shiri Harzevili}, \bibinfo{person}{Alvine~Boaye Belle}, \bibinfo{person}{Junjie Wang}, \bibinfo{person}{Song Wang}, \bibinfo{person}{Zhen Ming}, \bibinfo{person}{Nachiappan Nagappan}, {et~al\mbox{.}}} \bibinfo{year}{2023}\natexlab{}.
\newblock \showarticletitle{A survey on automated software vulnerability detection using machine learning and deep learning}.
\newblock \bibinfo{journal}{\emph{arXiv preprint arXiv:2306.11673}} (\bibinfo{year}{2023}).
\newblock


\bibitem[He and Vechev(2023)]%
        {he2023large}
\bibfield{author}{\bibinfo{person}{Jingxuan He} {and} \bibinfo{person}{Martin Vechev}.} \bibinfo{year}{2023}\natexlab{}.
\newblock \showarticletitle{Large language models for code: Security hardening and adversarial testing}. In \bibinfo{booktitle}{\emph{Proceedings of the 2023 ACM SIGSAC Conference on Computer and Communications Security}}. \bibinfo{pages}{1865--1879}.
\newblock


\bibitem[Hu et~al\mbox{.}(2015)]%
        {hu2015automatic}
\bibfield{author}{\bibinfo{person}{Hong Hu}, \bibinfo{person}{Zheng~Leong Chua}, \bibinfo{person}{Sendroiu Adrian}, \bibinfo{person}{Prateek Saxena}, {and} \bibinfo{person}{Zhenkai Liang}.} \bibinfo{year}{2015}\natexlab{}.
\newblock \showarticletitle{Automatic Generation of $\{$Data-Oriented$\}$ Exploits}. In \bibinfo{booktitle}{\emph{24th USENIX Security Symposium (USENIX Security 15)}}. \bibinfo{pages}{177--192}.
\newblock


\bibitem[Huang et~al\mbox{.}(2013)]%
        {huang2013craxweb}
\bibfield{author}{\bibinfo{person}{Shih-Kun Huang}, \bibinfo{person}{Han-Lin Lu}, \bibinfo{person}{Wai-Meng Leong}, {and} \bibinfo{person}{Huan Liu}.} \bibinfo{year}{2013}\natexlab{}.
\newblock \showarticletitle{Craxweb: Automatic web application testing and attack generation}. In \bibinfo{booktitle}{\emph{2013 IEEE 7th International Conference on Software Security and Reliability}}. IEEE, \bibinfo{pages}{208--217}.
\newblock


\bibitem[Jimenez et~al\mbox{.}(2023)]%
        {jimenez2023swe}
\bibfield{author}{\bibinfo{person}{Carlos~E Jimenez}, \bibinfo{person}{John Yang}, \bibinfo{person}{Alexander Wettig}, \bibinfo{person}{Shunyu Yao}, \bibinfo{person}{Kexin Pei}, \bibinfo{person}{Ofir Press}, {and} \bibinfo{person}{Karthik Narasimhan}.} \bibinfo{year}{2023}\natexlab{}.
\newblock \showarticletitle{Swe-bench: Can language models resolve real-world github issues?}
\newblock \bibinfo{journal}{\emph{arXiv preprint arXiv:2310.06770}} (\bibinfo{year}{2023}).
\newblock


\bibitem[Kang et~al\mbox{.}(2024)]%
        {kang2024evaluating}
\bibfield{author}{\bibinfo{person}{Sungmin Kang}, \bibinfo{person}{Juyeon Yoon}, \bibinfo{person}{Nargiz Askarbekkyzy}, {and} \bibinfo{person}{Shin Yoo}.} \bibinfo{year}{2024}\natexlab{}.
\newblock \showarticletitle{Evaluating diverse large language models for automatic and general bug reproduction}.
\newblock \bibinfo{journal}{\emph{IEEE Transactions on Software Engineering}} (\bibinfo{year}{2024}).
\newblock


\bibitem[Li et~al\mbox{.}(2025)]%
        {li2025iris}
\bibfield{author}{\bibinfo{person}{Ziyang Li}, \bibinfo{person}{Saikat Dutta}, {and} \bibinfo{person}{Mayur Naik}.} \bibinfo{year}{2025}\natexlab{}.
\newblock \showarticletitle{IRIS: LLM-assisted static analysis for detecting security vulnerabilities}. In \bibinfo{booktitle}{\emph{The Thirteenth International Conference on Learning Representations}}.
\newblock


\bibitem[Mei et~al\mbox{.}(2024)]%
        {mei2024arvo}
\bibfield{author}{\bibinfo{person}{Xiang Mei}, \bibinfo{person}{Pulkit~Singh Singaria}, \bibinfo{person}{Jordi Del~Castillo}, \bibinfo{person}{Haoran Xi}, \bibinfo{person}{Tiffany Bao}, \bibinfo{person}{Ruoyu Wang}, \bibinfo{person}{Yan Shoshitaishvili}, \bibinfo{person}{Adam Doup{\'e}}, \bibinfo{person}{Hammond Pearce}, \bibinfo{person}{Brendan Dolan-Gavitt}, {et~al\mbox{.}}} \bibinfo{year}{2024}\natexlab{}.
\newblock \showarticletitle{ARVO: Atlas of Reproducible Vulnerabilities for Open Source Software}.
\newblock \bibinfo{journal}{\emph{arXiv preprint arXiv:2408.02153}} (\bibinfo{year}{2024}).
\newblock


\bibitem[{MITRE Corporation}(2025)]%
        {cwe_mitre}
\bibfield{author}{\bibinfo{person}{{MITRE Corporation}}.} \bibinfo{year}{2025}\natexlab{}.
\newblock \bibinfo{title}{Common Weakness Enumeration}.
\newblock
\urldef\tempurl%
\url{https://cwe.mitre.org}
\showURL{%
\tempurl}
\newblock
\shownote{Accessed: July 18, 2025}.


\bibitem[Mu et~al\mbox{.}(2018)]%
        {mu2018understanding}
\bibfield{author}{\bibinfo{person}{Dongliang Mu}, \bibinfo{person}{Alejandro Cuevas}, \bibinfo{person}{Limin Yang}, \bibinfo{person}{Hang Hu}, \bibinfo{person}{Xinyu Xing}, \bibinfo{person}{Bing Mao}, {and} \bibinfo{person}{Gang Wang}.} \bibinfo{year}{2018}\natexlab{}.
\newblock \showarticletitle{Understanding the reproducibility of crowd-reported security vulnerabilities}. In \bibinfo{booktitle}{\emph{27th USENIX Security Symposium (USENIX Security 18)}}. \bibinfo{pages}{919--936}.
\newblock


\bibitem[{National Institute of Standards and Technology}(2025)]%
        {nist_nvd_2024}
\bibfield{author}{\bibinfo{person}{{National Institute of Standards and Technology}}.} \bibinfo{year}{2025}\natexlab{}.
\newblock \bibinfo{title}{National Vulnerability Database}.
\newblock
\urldef\tempurl%
\url{https://nvd.nist.gov/vuln}
\showURL{%
\tempurl}
\newblock
\shownote{Accessed: July 15, 2025}.


\bibitem[Neubig and Wang(2024)]%
        {neubig2024-openhands-codeact-2.1:}
\bibfield{author}{\bibinfo{person}{Graham Neubig} {and} \bibinfo{person}{Xingyao Wang}.} \bibinfo{year}{2024}\natexlab{}.
\newblock \showarticletitle{OpenHands CodeAct 2.1: An Open, State-of-the-Art Software Development Agent}.
\newblock \bibinfo{journal}{\emph{All Hands AI Blog}} (\bibinfo{date}{1 November} \bibinfo{year}{2024}).
\newblock
\urldef\tempurl%
\url{https://www.all-hands.dev/blog/openhands-codeact-21-an-open-state-of-the-art-software-development-agent}
\showURL{%
\tempurl}


\bibitem[Nikitopoulos et~al\mbox{.}(2021)]%
        {nikitopoulos2021crossvul}
\bibfield{author}{\bibinfo{person}{Georgios Nikitopoulos}, \bibinfo{person}{Konstantina Dritsa}, \bibinfo{person}{Panos Louridas}, {and} \bibinfo{person}{Dimitris Mitropoulos}.} \bibinfo{year}{2021}\natexlab{}.
\newblock \showarticletitle{CrossVul: a cross-language vulnerability dataset with commit data}. In \bibinfo{booktitle}{\emph{Proceedings of the 29th ACM Joint Meeting on European Software Engineering Conference and Symposium on the Foundations of Software Engineering}}. \bibinfo{pages}{1565--1569}.
\newblock


\bibitem[Serebryany(2017)]%
        {serebryany2017oss}
\bibfield{author}{\bibinfo{person}{Kostya Serebryany}.} \bibinfo{year}{2017}\natexlab{}.
\newblock \showarticletitle{$\{$OSS-Fuzz$\}$-Google's continuous fuzzing service for open source software}.
\newblock  (\bibinfo{year}{2017}).
\newblock


\bibitem[Simsek et~al\mbox{.}(2025)]%
        {simsek2025pocgen}
\bibfield{author}{\bibinfo{person}{Deniz Simsek}, \bibinfo{person}{Aryaz Eghbali}, {and} \bibinfo{person}{Michael Pradel}.} \bibinfo{year}{2025}\natexlab{}.
\newblock \showarticletitle{PoCGen: Generating Proof-of-Concept Exploits for Vulnerabilities in Npm Packages}.
\newblock \bibinfo{journal}{\emph{arXiv preprint arXiv:2506.04962}} (\bibinfo{year}{2025}).
\newblock


\bibitem[Wang et~al\mbox{.}(2024a)]%
        {wang2024executable}
\bibfield{author}{\bibinfo{person}{Xingyao Wang}, \bibinfo{person}{Yangyi Chen}, \bibinfo{person}{Lifan Yuan}, \bibinfo{person}{Yizhe Zhang}, \bibinfo{person}{Yunzhu Li}, \bibinfo{person}{Hao Peng}, {and} \bibinfo{person}{Heng Ji}.} \bibinfo{year}{2024}\natexlab{a}.
\newblock \showarticletitle{Executable code actions elicit better llm agents}. In \bibinfo{booktitle}{\emph{Forty-first International Conference on Machine Learning}}.
\newblock


\bibitem[Wang et~al\mbox{.}(2024b)]%
        {neubig2024-opendevin-codeact-1.0:}
\bibfield{author}{\bibinfo{person}{Xingyao Wang}, \bibinfo{person}{Bowen Li}, {and} \bibinfo{person}{Graham Neubig}.} \bibinfo{year}{2024}\natexlab{b}.
\newblock \showarticletitle{Introducing OpenDevin CodeAct 1.0, a new State-of-the-art in Coding Agents}.
\newblock \bibinfo{journal}{\emph{Blog}} (\bibinfo{date}{7 May} \bibinfo{year}{2024}).
\newblock
\urldef\tempurl%
\url{https://xwang.dev/blog/2024/opendevin-codeact-1.0-swebench/}
\showURL{%
\tempurl}


\bibitem[Wang et~al\mbox{.}(2025)]%
        {wang2025openhands}
\bibfield{author}{\bibinfo{person}{Xingyao Wang}, \bibinfo{person}{Boxuan Li}, \bibinfo{person}{Yufan Song}, \bibinfo{person}{Frank~F. Xu}, \bibinfo{person}{Xiangru Tang}, \bibinfo{person}{Mingchen Zhuge}, \bibinfo{person}{Jiayi Pan}, \bibinfo{person}{Yueqi Song}, \bibinfo{person}{Bowen Li}, \bibinfo{person}{Jaskirat Singh}, \bibinfo{person}{Hoang~H. Tran}, \bibinfo{person}{Fuqiang Li}, \bibinfo{person}{Ren Ma}, \bibinfo{person}{Mingzhang Zheng}, \bibinfo{person}{Bill Qian}, \bibinfo{person}{Yanjun Shao}, \bibinfo{person}{Niklas Muennighoff}, \bibinfo{person}{Yizhe Zhang}, \bibinfo{person}{Binyuan Hui}, \bibinfo{person}{Junyang Lin}, \bibinfo{person}{Robert Brennan}, \bibinfo{person}{Hao Peng}, \bibinfo{person}{Heng Ji}, {and} \bibinfo{person}{Graham Neubig}.} \bibinfo{year}{2025}\natexlab{}.
\newblock \showarticletitle{OpenHands: An Open Platform for {AI} Software Developers as Generalist Agents}. In \bibinfo{booktitle}{\emph{The Thirteenth International Conference on Learning Representations}}.
\newblock
\urldef\tempurl%
\url{https://openreview.net/forum?id=OJd3ayDDoF}
\showURL{%
\tempurl}


\bibitem[Yang et~al\mbox{.}(2024)]%
        {yang2024swe}
\bibfield{author}{\bibinfo{person}{John Yang}, \bibinfo{person}{Carlos~E Jimenez}, \bibinfo{person}{Alexander Wettig}, \bibinfo{person}{Kilian Lieret}, \bibinfo{person}{Shunyu Yao}, \bibinfo{person}{Karthik Narasimhan}, {and} \bibinfo{person}{Ofir Press}.} \bibinfo{year}{2024}\natexlab{}.
\newblock \showarticletitle{Swe-agent: Agent-computer interfaces enable automated software engineering}.
\newblock \bibinfo{journal}{\emph{Advances in Neural Information Processing Systems}}  \bibinfo{volume}{37} (\bibinfo{year}{2024}), \bibinfo{pages}{50528--50652}.
\newblock


\bibitem[You et~al\mbox{.}(2017)]%
        {you2017semfuzz}
\bibfield{author}{\bibinfo{person}{Wei You}, \bibinfo{person}{Peiyuan Zong}, \bibinfo{person}{Kai Chen}, \bibinfo{person}{XiaoFeng Wang}, \bibinfo{person}{Xiaojing Liao}, \bibinfo{person}{Pan Bian}, {and} \bibinfo{person}{Bin Liang}.} \bibinfo{year}{2017}\natexlab{}.
\newblock \showarticletitle{Semfuzz: Semantics-based automatic generation of proof-of-concept exploits}. In \bibinfo{booktitle}{\emph{Proceedings of the 2017 ACM SIGSAC conference on computer and communications security}}. \bibinfo{pages}{2139--2154}.
\newblock


\bibitem[Zhang et~al\mbox{.}(2024)]%
        {zhang2024autocoderover}
\bibfield{author}{\bibinfo{person}{Yuntong Zhang}, \bibinfo{person}{Haifeng Ruan}, \bibinfo{person}{Zhiyu Fan}, {and} \bibinfo{person}{Abhik Roychoudhury}.} \bibinfo{year}{2024}\natexlab{}.
\newblock \showarticletitle{Autocoderover: Autonomous program improvement}. In \bibinfo{booktitle}{\emph{Proceedings of the 33rd ACM SIGSOFT International Symposium on Software Testing and Analysis}}. \bibinfo{pages}{1592--1604}.
\newblock


\bibitem[Zhu et~al\mbox{.}(2025)]%
        {zhu2025cve}
\bibfield{author}{\bibinfo{person}{Yuxuan Zhu}, \bibinfo{person}{Antony Kellermann}, \bibinfo{person}{Dylan Bowman}, \bibinfo{person}{Philip Li}, \bibinfo{person}{Akul Gupta}, \bibinfo{person}{Adarsh Danda}, \bibinfo{person}{Richard Fang}, \bibinfo{person}{Conner Jensen}, \bibinfo{person}{Eric Ihli}, \bibinfo{person}{Jason Benn}, {et~al\mbox{.}}} \bibinfo{year}{2025}\natexlab{}.
\newblock \showarticletitle{CVE-Bench: A Benchmark for AI Agents' Ability to Exploit Real-World Web Application Vulnerabilities}.
\newblock \bibinfo{journal}{\emph{arXiv preprint arXiv:2503.17332}} (\bibinfo{year}{2025}).
\newblock


\bibitem[Zhu et~al\mbox{.}(2024)]%
        {zhu2024teams}
\bibfield{author}{\bibinfo{person}{Yuxuan Zhu}, \bibinfo{person}{Antony Kellermann}, \bibinfo{person}{Akul Gupta}, \bibinfo{person}{Philip Li}, \bibinfo{person}{Richard Fang}, \bibinfo{person}{Rohan Bindu}, {and} \bibinfo{person}{Daniel Kang}.} \bibinfo{year}{2024}\natexlab{}.
\newblock \showarticletitle{Teams of llm agents can exploit zero-day vulnerabilities}.
\newblock \bibinfo{journal}{\emph{arXiv preprint arXiv:2406.01637}} (\bibinfo{year}{2024}).
\newblock


\end{thebibliography}
